**Stacking-orientation and twist-angle control on integer and fractional Chern insulators in moiré rhombohedral graphene**

Chushan Li[1,2†], Chuanqi Zheng[1†], Kai Liu[1,2], Ke Huang[3], Zheng Sun[2], Lei Qiao[4,5], Yifan Wei[2], Chenyu Zhang[1], Fan Xu[2], Kenji Watanabe[6], Takashi Taniguchi[7], Hao Yang[1,2,8], Dandan Guan[1,2,8], Liang Liu[1,2,8], Shiyong Wang[1,2,8], Yaoyi Li[1,2,8], Hao Zheng[1,2,8], Canhua Liu[1,2,8], Bingbing Tong[9], Li Lu[8,9], Jinfeng Jia[1,2,8,10], Zhiwen Shi[1,2], Jianpeng Liu[5,11], Xiao Li[3], Guorui Chen[1,2*], Tingxin Li[1,2,8*] and Xiaoxue Liu[1,2,8*]

[1] Tsung-Dao Lee Institute, Shanghai Jiao Tong University, Shanghai,201210, China

[2] Key Laboratory of Artificial Structures and Quantum Control (Ministry of Education), School of Physics and Astronomy, Shanghai Jiao Tong University, Shanghai 200240, China

[3] Department of Physics, City University of Hong Kong, Kowloon, Hong Kong SAR, China

[4] Kavli Institute for Theoretical Sciences, University of Chinese Academy of Sciences, Beijing 100190, China

[5] School of Physical Science and Technology, ShanghaiTech Laboratory for Topological Physics, ShanghaiTech University, Shanghai 201210, China

[6] Research Center for Electronic and Optical Materials, National Institute for Materials Science, 1-1 Namiki, Tsukuba, Japan

[7] Research Center for Materials Nanoarchitectonics, National Institute for Materials Science, 1-1 Namiki, Tsukuba, Japan

[8] Hefei National Laboratory, Hefei, China

[9] Beijing National Laboratory for Condensed Matter Physics and Institute of Physics, Chinese Academy of Sciences, Beijing, China

[10] Department of Physics, Southern University of Science and Technology, Shenzhen 518055, China

[11] Liaoning Academy of Materials, Shenyang 110167, China

[†]These authors contributed equally to this work.

[*]Emails: chenguorui@sjtu.edu.cn, txli89@sjtu.edu.cn, xxliu90@sjtu.edu.cn

## Abstract

Rhombohedral-stacked multilayer graphene (RMG) aligned with hexagonal boron nitride (hBN) has emerged as an excellent platform for investigating exotic quantum phenomena arising from the interplay between electron correlations and nontrivial topology. However, the microscopic mechanism governing the emergence of both the integer and fractional Chern insulator states in this system remains an open question. In this work, we systematically investigate the electrical transport properties of RMG/hBN moiré devices with controlled alignment orientations and twist angles. We demonstrate that alignment orientation strongly modulates correlated phenomena in the moiré-proximal regime, while having negligible influence on the formation of integer and fractional Chern insulators in the moiré-distant regime. Instead, the moiré periodicity—tuned by the twist angle—serves as the key parameter controlling the stability of these correlated topological states in the moiré-distant regime. Furthermore, in the moiré-proximal regime of one specific alignment, we observe anomalous Hall effect and a variety of competing phases near $\nu = 1$, including integer Chern insulator states, extended Chern insulator states, and trivial insulators, whose stability is highly sensitive to both the applied displacement electric field and magnetic field. Our results underscore the critical role of stacking-alignment and twist-angle engineering in exploring novel quantum states based on rhombohedral-stacked multilayer graphene moiré systems.

## Introduction

Recently, based on rhombohedral-stacked multilayer graphene (RMG) aligned with hexagonal boron nitride (hBN) moiré systems, a rich variety of quantum phases have been uncovered, including correlated insulators [1-3], superconductivity [4,5], orbital ferromagnetism and Chern insulators [6-17], among others. Notably, the zero-field fractional Chern insulator (FCI) [18-23], namely the fractional quantum anomalous Hall (FQAH) effect [24,25], has been observed in tetralayer [5,12], pentalayer [8,12], and hexalayer [11] RMG aligned with hBN moiré superlattices. Strikingly, such novel correlated topological states only manifest in the moiré-distant regime where the electrons are pushed away from the RMG/hBN moiré interface by the applied displacement electric field *D*. More generally, in 3- to 10-layer RMG aligned with hBN systems, interaction-induced orbital ferromagnetism and quantum anomalous Hall effects have always been observed in the moiré-distant regime [5-17]. In contrast, in the moiré-proximal regime (where carriers are polarized close to the RMG/hBN moiré interface by the applied *D*-field), only trivial correlated insulating states were observed, except a recent compressibility study [14] has revealed the coexistence of integer and fractional Chern insulators, as well as topologically trivial and nontrivial charge density waves (CDW) states under finite magnetic fields. The underlying role of the moiré superlattice potential in developing the moiré Chern band and correlated topological phases remains under debate [26-39]. In fact, the effective moiré potential is not only controlled by the moiré lattice periodicity (i.e. the twist angle between hBN and RMG), but also influenced by the specific hBN/RMG stacking configurations [40]. A systematic experimental investigation of the effects of both the twist angle and the hBN stacking orientation on the correlated topological states in RMG/hBN moiré system is still lacking.

In this study, we present electrical transport studies of rhombohedral tetralayer graphene (RTG) and rhombohedral pentalayer graphene (RPG) aligned with hBN at various twist angles, and with two distinct stacking orientations. We found that in the moiré-distant regime, the existence of both integer and fractional Chern insulators is insensitive to the alignment between RMG and hBN, but depends on the moiré periodicity. In stark contrast, in the moiré-proximal regime, the alignment orientation of RMG/hBN significantly influences the formation of correlated and topological phases. For one alignment, only trivial correlated insulators are observed, whereas under the other alignment, Chern insulators emerge under a small magnetic field, accompanied by orbital ferromagnetism at zero magnetic field.

## Global transport phase diagram with opposite stacking orientations

Stacking orientations between moiré layers can profoundly influence band topology and correlated ground state, as demonstrated in previous studies on $MoTe_2/WSe_2$ semiconductor moiré systems [41,42]. In the RMG/hBN moiré system, due to the breaking of $C_{2z}$ symmetry in both hBN and RMG, there exist two distinct stacking configurations, commonly referred to as $\xi = 0$ and $\xi = 1$ [32,40]. To systematically investigate the effect of hBN alignment orientation on the emergent correlated and topological phases in RMG, we fabricated two RTG/hBN devices (Fig. 1a) with similar moiré periodicities but opposite alignment orientations between RTG and hBN. Figure 1b presents a schematic illustration of the device stacking procedure. In short, we cut a piece of hBN and a piece of RTG each in half, and fabricated them into two separate devices with targeted alignment angles of 0° and 180°, respectively (See Methods and Extended Data Fig. 1 for detailed device fabrication procedures). By doing so, we obtained a controlled pair for comparative analysis of the stacking-orientation effect in this system, although conclusively determining the atomic stacking configuration in each device remains challenging. Figures 1c and 1d show the longitudinal resistivity $\rho_{xx}$ as a function of moiré filling factors $v$ and $D$ for these two devices (device 1 and device 2), measured at 1.7 K. The moiré periodicities of device 1 and device 2 are approximately 14.1 nm and 13.2 nm, respectively, as calibrated from the Brown-Zak oscillations (see Extended Data Fig. 2).

At the charge neutrality point, $\rho_{xx}$ as a function of $D$ for both devices exhibit behaviors similar to those observed in moiréless RMG [43,44], with the system transitioning from a layer antiferromagnetic insulator to a semimetal, and eventually to a layer-polarized insulator at high $D$. When carriers are doped away from the charge neutrality point, moiré-induced correlated insulating states or Chern insulators emerge within certain ranges of $D$ at commensurate moiré fillings (e.g. $v = \pm 1, \pm 2, 3$), on both the electron- and hole-doped sides, as previously reported [5,12]. In this study, we focus on the regime of high displacement fields on the electron-doped side, where the moiré-distant and moiré-proximal regime are more well defined. In all devices presented in the following, a positive (negative) $D$ corresponds to the moiré-distant (moiré-proximal) regime for electron doping.

First of all, we notice that the hBN alignment orientation has minimal influence on the transport behavior in the moiré-distant region, where the overall features are quite consistent across both device 1 and device 2, as shown in Figs. 1c and 1d, respectively. This is consistent with the scenario where the electronic wavefunction is polarized away from the moiré interface, resulting in a relatively weak moiré potential that is less sensitive to specific moiré stacking configurations. In contrast, the moiré-proximal region on the electron-doped side exhibit distinct characteristics between the two devices. For device 1 shown in Fig. 1c, a cascade of insulating states emerges at $v = 1, 2, 3$ and 4 within a broad range of $D$, with the measured $\rho_{xx}$ at 1.7 K exceeding 1 Mohm. In device 2, although enhanced $\rho_{xx}$ is also observed at integer moiré filling factors, it remains metallic with the resistance on the order of several to tens of kilo-ohms except at $v = 1$ and 2. And the overall shape of the enhanced $\rho_{xx}$ features closely resembles that of the moiré-distant case. A recent study [40] pointed out that strong insulating behavior at $v = 3$ and 4 on the moiré-proximal regime serves as a robust empirical signature of $\xi = 1$ devices. We adopt the same criterion here in this paper.

Although this pair of devices (device 1 and device 2) have slightly different moiré periodicities, we emphasize that the distinct transport behaviors arise primarily from their opposite hBN alignment orientations. Firstly, the difference in moiré periodicities between the two devices is small—within 7%. In addition, as shown in the following sections, measurement results from other devices demonstrate that devices with very similar moiré periodicities can exhibit strikingly different transport behaviors in the moiré-proximal region. Taken together, these observations suggest that the differences in the moiré-proximal regime are primarily due to the opposite hBN alignment orientations, rather than variations in the twist angle.

**IQAH and FQAH effects in the moiré-distant regime**

We further examine the influence of hBN alignment orientation on the emergence of both the integer quantum anomalous Hall (IQAH) effect and the FQAH effect in the moiré distant regime. Figure 2a and 2h show the $\rho_{xx}$ as a function of $\nu$ and $D$ on the electron-doped side, measured in two RTG/hBN devices with $\xi = 1$ alignment (device 3) and $\xi = 0$ alignment (device 2), respectively. In particular, device 2 and device 3 have very similar moiré periodicity, 13.2 nm and 13.0 nm, with a difference of less than 2%.

In the moiré-distant region, the symmetrized $\rho_{xx}$ and antisymmetrized $\rho_{xy}$ of device 3 (device 2) as functions of $\nu$ and $D$, measured at $B_{\perp} = \pm$ 0.1 T, are displayed in Figs. 2b and 2c (Figs. 2i and 2j), respectively. Consistent with previous studies on RMG/hBN systems [8,11-13], large anomalous Hall (AH) signals have been observed in a tilted region in the $\nu$-$D$ map. Figure 2d (2k) shows the $\rho_{xx}$ and $\rho_{xy}$ along the dashed line shown in Figs. 2b and 2c (Figs. 2i and 2j). Clearly, in both devices, the IQAH effect is observed at $\nu = 1$, and FQAH effects appear at fractional fillings $\nu = 2/3$, 3/5, and 4/7, evidenced by quantized $\rho_{xy}$ plateaus along with $\rho_{xx}$ minimums with tuning the electron density (see Extended Data Fig. 3 for the temperature dependence). Moreover, in the Landau fan diagrams (Extended Data Fig. 3), $\rho_{xx}$ and $\rho_{xy}$ at both the IQAH and FQAH states disperse with perpendicular magnetic field $B_{\perp}$ and carrier density following Streda formula with expected Chern numbers. Magnetic hysteresis loops of IQAH and FQAH effects are shown in Figs 2e–2g (Figs. 2l-2n) for device 3 (device 2). Overall, these experimental observations unambiguously demonstrate that both the IQAH and FQAH effects in the moiré-distant regime are robust against the two distinct stacking orientations between RTG and hBN. This is inconsistent with single-particle and Hartree-Fock (HF) calculation results [32,45], which show that the moiré band Chern number and the phase diagram in the moiré-distant regime depends on the alignment orientation between RMG and hBN. This discrepancy highlights the need for further investigations into the mechanism of the IQAH and FQAH effects in RMG moiré systems, beyond the mean-field approach.

We have also investigated the twist-angle dependence of the Chern insulator states. As summarized in Table I and Extended data Fig. 4, we find that as the moiré periodicity decreases, the $C = 1$ IQAH effect generally requires a larger $D$ for stabilization and exists over a narrowing range of $D$. In an RPG/hBN device with relatively short moiré periodicity (approximately 9.9 nm, device 7), AH signals are observed at zero magnetic field over a large range of $\nu$ in the moiré-distant region, but it is not quantized even at $\nu = 1$. It indicates the ground state at $\nu = 1$ and $B_{\perp} = 0$ in device 7 is an AH metal state rather than an IQAH insulator. Upon applying an external $B_{\perp}$, $\rho_{xx}$ starts developing a dip along with enhanced $\rho_{xy}$, and becomes quantized at $h/e^2$ as $B_\perp$ beyond 3 T (Extended Data Fig. 5). These observations demonstrate that the twist angle serve as an effective tuning knob for the IQAH states in RMG/hBN systems. As for the fractional filling states, we observe a more subtle evolution with the twist angles. FCI states have been observed in four RTG/hBN devices with twist angles below 0.66 degrees (Fig. 2, Extended Data Fig. 4). Additionally, in an RTG/hBN device (device 6) with a 0.78-degree twist angle, we observed extended quantum anomalous Hall states [12] over a continuous range of fractional filling factors (Extended Data Fig. 4). The evolution of FCI states and their competing phases with respect to twist angle and the layer number of rhombohedral-stacked graphene requires further in-depth theoretical and experimental investigation.

**Alignment-dependent phase diagram in the moiré-proximal regime**

We now discuss the alignment-orientation-dependent phase diagram in the moiré-proximal region. Figure 3a and 3d show $\rho_{xx}$ as a function of $v$ and $D$ in device 1 ($\xi = 1$ alignment) and device 4 ($\xi = 0$ alignment), respectively. The moiré periodicities of device 1 and device 4 are very similar—14.1 nm and 14.4 nm, respectively. Figure 3b presents the longitudinal conductivity $\sigma_{xx}$ as a function of $v$ at various temperatures in the moiré-proximal regime for device 4, with $D$ fixed at -0.61 V/nm. It is evident that in the moiré-proximal region with $\xi = 1$ alignment, the states at $v = 1$, 2, 3, and 4 all exhibit insulating behaviors up to 30 K (also see Extended Data Fig. 6). Considering both the spin and valley degeneracies, the insulating states at $v = 1$, 2, and 3 correspond to correlated insulators, while the $v = 4$ insulating states corresponds to a band insulator. In contrast to the moiré-distant region, no magnetic hysteresis loop and AH effect is observed in the moiré-proximal region for $\xi = 1$ devices. Representative examples are shown in Fig. 3c and Extended data Fig. 6, illustrating the absence of net magnetization or spontaneous time-reversal symmetry breaking phases.

As previously mentioned, the transport behaviors in the moiré-proximal region of devices with $\xi = 0$ alignment significantly differ from those with $\xi = 1$ alignment. Specifically, (i) the correlated insulators at $v = 1$ and 2 are much less insulating than those in the $\xi = 1$ devices; (ii) the correlated insulator at $v = 3$ is absent; and (iii) the $\rho_{xx}$ at the $v = 4$ is significantly lower—only on the order of ten kilo-ohms. Single-particle calculations also reveal that (Extended Data Fig. 7), for $\xi = 1$ aligned RTG/hBN, the lowest moiré conduction band in the moiré-proximal regime is an isolated flat band, while for $\xi = 0$ alignment, the lowest moiré conduction band overlaps with remote moiré bands. By performing Hartree-Fock (HF) calculations (Methods), we find that the interaction is capable of opening a correlation gap at $v = 1$ and 2 for both the $\xi = 1$ and $\xi = 0$ aligned RTG/hBN systems. Notwithstanding, the $\xi = 1$ case possesses a much larger gap than the $\xi = 0$ case, consistent with the experimental observations. Furthermore, the $\xi = 1$ system also forms insulating ground states at $v = 3$ and 4, while for the $\xi = 0$ alignment, the system favors a metallic ground state.

We further examine the transport behavior near $v = 1$ in the moiré-proximal region for devices with $\xi = 0$ alignment. Figure 3e presents the antisymmetrized $\rho_{xy}$ as a function of $v$ and $D$ at $B_\perp = \pm 0.1$ T of device 4. Figure 3f and 3g shows a representative magnetic hysteresis loop with AH effect near $v = 1$ (additional data are provided in Extended Data Fig. 8), illustrating the emergence of orbital ferromagnetism near $v = 1$ in the moiré-proximal region, which is distinct from the behavior observed in $\xi = 1$ aligned devices. It is important to note that such alignment-dependent phase diagram in the moiré-proximal region is not limited to RTG. Figure 4 show data measured from device 7 (RPG/hBN with $\xi = 0$ alignment), which exhibit similar behaviors in the moiré-proximal region as those observed in RTG. Figure 4b shows the anti-symmetrized map of $\rho_{xy}$ as functions of $D$ and $\nu$ for device 7 under a small magnetic field of $B_\perp = \pm 50$ mT. Again, we observe AH effects with clear magnetic hysteresis near $\nu = 1$ in this device, as shown in Fig. 4c and Extended Data Fig. 9.

## Rich competing phases near $\nu$ = 1 in the moiré-proximal regime

Previous studies have mainly shown that the conduction band of RMG/hBN system in the moiré-proximal regime is topologically trivial [5-13,15-17]. Here, we revealed a rich phase diagram in the moiré-proximal regime for devices with $\xi = 0$ alignment. As shown in Fig. 4a-4c, in the moiré-proximal region of device 7, the applied $D$-field induce a phase transition from an AH metal to a trivial correlated insulator at $\nu = 1$. Figures 4d and 4e show symmetrized $\rho_{xx}$ and antisymmetrized $\rho_{xy}$ as a function of $\nu$ and $B_\perp$ at $D = -0.72$ V/nm in the moiré-proximal region. At $\nu = 1$, when applying a small $B_\perp$ about 0.3 T, $\rho_{xy}$ reached the quantized value of $-h/e^2$ accompanied by the vanishing $\rho_{xx}$, as shown in Fig. 4f, corresponding to an integer Chern insulator (ICI) state with $C = -1$. As $B_\perp$ increases, a trivial correlated insulator begins to emerge at approximately $B_\perp = 0.6$ T, exhibiting an obvious curved feature in the $\nu$-$B_\perp$ map. The ICI state with $C = -1$ also follows the curved trajectory, surrounding the trivial correlated insulator. As further increasing $B_\perp$ up to 1.4 T, another ICI state with $C = +1$ emerges on the higher doping side with respect to the trivial insulator. We note that, these two ICI states with $C = +1$ and -1 extend over a broad moiré filling range. Specifically, the quantized Hall resistance of $\rho_{xy} = +h/e^2$ or $\rho_{xy} = -h/e^2$ with vanishing $\rho_{xx}$ is observed at $1.05 < \nu < 1.45$, and $0.8 < \nu < 1.35$, respectively. Similar extended Chern insulator states have been recently observed in the magic-angle twisted bilayer graphene aligned with hBN moiré system [46], where the quantized Hall resistance extends to a large carrier density range and deviates from the Streda formula. The underlying mechanism is believed to be associated with the formation of electronic crystals upon doping a Chern insulator [12,46]. When further increasing $B_\perp$ to 4 T, the trivial correlated insulator becomes more stabilized, while the ICI state with $C = -1$ is fully suppressed. Overall, these observations illustrate a competing nature between Chern insulator states and trivial correlated insulators in the moiré-proximal regime.

Aside from the $B_\perp$, $D$ also plays a crucial role in determining the ground states. Extended Data Fig. 10 shows the competitions between the trivial correlated insulator and the ICI states with $C = \pm 1$ near $\nu = 1$ at $B_\perp = 1$ T in the $\nu$-$D$ phase space for device 7. This once again demonstrates that various quantum phases in the moiré-proximal region may have comparable energy scales, and the ground state is highly sensitive to both the applied $B_\perp$ and $D$. Similar competitions between ICI states and trivial insulators near $\nu = 1$ have also been observed in the moiré-proximal regime of an RTG device with $\xi = 0$ alignment (device 4, see Extended Data Fig. 8).

## Conclusions

In conclusion, our results unambiguously demonstrate that, in the moiré-distant regime, both integer and fractional quantum anomalous Hall effects in rhombohedral-stacked graphene are robust against the hBN-RMG alignment orientation, yet sensitive to moiré periodicities. In the moiré-proximal regime, the phase diagram of emergent quantum states is primarily governed by the hBN-RMG alignment orientation. Specifically, for $\xi = 0$ aligned devices in the moiré-proximal regime, we observe competitions between integer Chern insulators, extended Chern insulator states and a trivial correlated insulator around $\nu = 1$. These findings are crucial for understanding the formation of the FQAH effect and other correlated topological phases in this system, emphasizing stacking alignment and twist angle as two central and complementary control parameters for engineering exotic quantum phases in rhombohedral graphene aligned with hBN moiré systems.

## References

[1] G. Chen *et al.*, *Evidence of a gate-tunable Mott insulator in a trilayer graphene moiré superlattice*, Nat. Phys. **15**, 237 (2019).

[2] H. Zhou *et al.*, *Half- and quarter-metals in rhombohedral trilayer graphene*, Nature (London) **598**, 429-433 (2021).

[3] X. Han *et al.*, *Suppression of symmetry-breaking correlated insulators in a rhombohedral trilayer graphene superlattice*, Nat. Commun. **15**, 9765 (2024).

[4] G. Chen *et al.*, *Signatures of tunable superconductivity in a trilayer graphene moiré superlattice*, Nature (London) **572**, 215 (2019).

[5] Y. Choi *et al.*, *Superconductivity and quantized anomalous Hall effect in rhombohedral graphene*, Nature (London) **639**, 342-347 (2025).

[6] G. Chen *et al.*, *Tunable correlated Chern insulator and ferromagnetism in a moiré superlattice*, Nature (London) **579**, 56 (2020).

[7] G. Chen *et al.*, *Tunable Orbital Ferromagnetism at Noninteger Filling of a Moiré Superlattice*, Nano Lett. **22**, 238 (2022).

[8] Z. Lu *et al.*, *Fractional quantum anomalous Hall effect in multilayer graphene*, Nature (London) **626**, 759 (2024).

[9] S. Wang *et al.*, *Chern Insulator States with Tunable Chern Numbers in a Graphene Moiré Superlattice*, Nano Lett. **24**, 6838 (2024).

[10] X. Han *et al.*, *Engineering the Band Topology in a Rhombohedral Trilayer Graphene Moiré Superlattice*, Nano Lett. **24**, 6286 (2024).

[11] J. Xie *et al.*, *Tunable fractional Chern insulators in rhombohedral graphene superlattices*, Nature Materials (2025).

[12] Z. Lu *et al.*, *Extended quantum anomalous Hall states in graphene/hBN moiré superlattices*, Nature (London) **637**, 1090 (2025).

[13] D. Waters *et al.*, *Chern Insulators at Integer and Fractional Filling in Moiré Pentalayer Graphene*, Phys. Rev. X **15**, 011045 (2025).

[14] S. H. Aronson, T. Han, Z. Lu, Y. Yao, K. Watanabe, T. Taniguchi, L. Ju, and R. C. Ashoori, *Displacement field-controlled fractional Chern insulators and charge density waves in a graphene/hBN moiré superlattice*, Phys. Rev. X **15**, 031026 (2025).

[15] J. Ding *et al.*, *Electric-Field Switchable Chirality in Rhombohedral Graphene Chern Insulators Stabilized by Tungsten Diselenide*, Phys. Rev. X **15**, 011052 (2025).

[16] J. Zheng *et al.*, *Switchable Chern insulators and competing quantum phases in rhombohedral graphene moiré superlattices*, Phys. Rev. Lett. **135**, 136302 (2025).

[17] Z. Wang *et al.*, *Electrical switching of Chern insulators in moiré rhombohedral heptalayer graphene*, arXiv:2503.00837.

[18] F. D. M. Haldane, *Model for a Quantum Hall Effect without Landau Levels: Condensed-Matter Realization of the "Parity Anomaly"*, Phys. Rev. Lett. **61**, 2015 (1988).

[19] E. Tang, J.-W. Mei, and X.-G. Wen, *High-Temperature Fractional Quantum Hall States*, Phys. Rev. Lett. **106**, 236802 (2011).

[20] K. Sun, Z. Gu, H. Katsura, and S. Das Sarma, *Nearly Flatbands with Nontrivial Topology*, Phys. Rev. Lett. **106**, 236803 (2011).

[21] T. Neupert, L. Santos, C. Chamon, and C. Mudry, *Fractional Quantum Hall States at Zero Magnetic Field*, Phys. Rev. Lett. **106**, 236804 (2011).

[22] N. Regnault and B. A. Bernevig, *Fractional Chern Insulator*, Phys. Rev. X **1**, 021014 (2011).

[23] D. N. Sheng, Z.-C. Gu, K. Sun, and L. Sheng, *Fractional quantum Hall effect in the absence of Landau levels*, Nat. Commun. **2**, 389 (2011).

[24] H. Park *et al.*, *Observation of fractionally quantized anomalous Hall effect*, Nature (London) **622**, 74 (2023).

[25] F. Xu *et al.*, *Observation of Integer and Fractional Quantum Anomalous Hall Effects in Twisted Bilayer $MoTe_2$*, Phys. Rev. X **13**, 031037 (2023).

[26] A. S. Patri, T. Senthil, *Strong correlations in ABC-stacked trilayer graphene: Moiré is important*, Phys. Rev. B **107**, 165122 (2023).

[27] Z. Dong, A. S. Patri, and T. Senthil, *Theory of Quantum Anomalous Hall Phases in Pentalayer Rhombohedral Graphene Moiré Structures*, Phys. Rev. Lett. **133**, 206502 (2024).

[28] J. Dong, T. Wang, T. Wang, T. Soejima, M. P. Zaletel, A. Vishwanath, and D. E. Parker, *Anomalous Hall Crystals in Rhombohedral Multilayer Graphene. I. Interaction-Driven Chern Bands and Fractional Quantum Hall States at Zero Magnetic Field*, Phys. Rev. Lett. **133**, 206503 (2024).

[29] B. Zhou, H. Yang, and Y. H. Zhang, *Fractional quantum anomalous Hall effects in rhombohedral multilayer graphene in the moiréless limit and in Coulomb imprinted superlattice*, Phys. Rev. Lett. **133**, 206504 (2024).

[30] T. Soejima *et al.*, *Anomalous Hall crystals in rhombohedral multilayer graphene. II: General mechanism and a minimal model*, Phys. Rev. B **110**, 205124 (2024).

[31] J. Herzog-Arbeitman *et al.*, *Moiré fractional Chern insulators. II. First-principles calculations and continuum models of rhombohedral graphene superlattices*, Phys. Rev. B **109**, 205122 (2024).

[32] Y. H. Kwan *et al.*, *Moiré Fractional Chern Insulators III: Hartree-Fock Phase Diagram, Magic Angle Regime for Chern Insulator States, the Role of the Moiré Potential and Goldstone Gaps in Rhombohedral Graphene Superlattices*, Phys. Rev. B **112**, 075109 (2025).

[33] Yu, J., Herzog-Arbeitman, J., Kwan, Y. H., Regnault, N. & Bernevig, B. A. *Moiré fractional Chern insulators. IV. Fluctuation-driven collapse in multiband exact diagonalization calculations on rhombohedral graphene.* Phys. Rev. B 112, 075110 (2025).

[34] Tan, T. & Devakul, T. *Parent Berry Curvature and the Ideal Anomalous Hall Crystal*. Phys. Rev. X 14, 041040 (2024).

[35] Z. Guo, X. Lu, B. Xie, and J. Liu, *Fractional Chern insulator states in multilayer graphene moiré superlattices*, Phys. Rev. B **110**, 075109 (2024).

[36] K. Huang, X. Li, S. Das Sarma, and F. Zhang, *Self-consistent theory of fractional quantum anomalous Hall states in rhombohedral graphene*, Phys. Rev. B **110**, 115146 (2024).

[37] K. Huang, S. Das Sarma, and X. Li, *Fractional quantum anomalous Hall effect in rhombohedral multilayer graphene with a strong displacement field*, Phys. Rev. B **111**, 075130 (2025).

[38] Z. Dong, A. S. Patri, and T. Senthil, *Stability of anomalous Hall crystals in multilayer rhombohedral graphene*, Phys. Rev. B **110**, 205130 (2024).

[39] V. Crépel and J. Cano. Efficient Prediction of Superlattice and Anomalous Miniband Topology from Quantum Geometry. Phys. Rev. X **15**, 011004 (2025).

[40] M. Uzan *et al.*, *hBN alignment orientation controls moire strength in rhombohedral graphene.* arXiv:2507.20647.

[41] T. Li *et al.*, *Continuous Mott transition in semiconductor moiré superlattices.*, Nature (London) **597**, 350 (2021).

[42] T. Li *et al.*, *Quantum anomalous Hall effect from intertwined moiré bands*, Nature (London) **600**, 641 (2021).

[43] T. Han *et al.*, *Correlated insulator and Chern insulators in pentalayer rhombohedral-stacked graphene*, Nat. Nanotechnol. **19**, 181 (2024).

[44] K. Liu *et al.*, *Spontaneous broken-symmetry insulator and metals in tetralayer rhombohedral graphene*, Nat. Nanotechnol. **19**, 188 (2024).

[45] Y. Park, Y. Kim, B. L. Chittari, and J. Jung. *Topological flat bands in rhombohedral tetralayer and multilayer graphene on hexagonal boron nitride moiré superlattices*. Phys. Rev. B **108**, 155406 (2023).

[46] Z. Zhang *et al.*, *Cascade of Zero-field Chern Insulators in Magic-angle Bilayer Graphene*, National Science Review, https://doi.org/10.1093/nsr/nwaf265

[47] Moon, P. & Koshino, M. *Electronic properties of graphene/hexagonal-boron-nitride moiré superlattice*, Phys. Rev. B **90**, 155406 (2014).

[48] Jung, J., Raoux, A., Qiao, Z. & MacDonald, A. H. *Ab initio theory of moiré superlattice bands in layered two-dimensional materials*, Phys. Rev. B **89**, 205414 (2014).

[49] Zhang, F., Sahu, B., Min, H. & MacDonald, A. H. *Band structure of ABC -stacked graphene trilayers*, Phys. Rev. B **82**, 035409 (2010).

[50] Jung, J. & MacDonald, A. H. *Gapped broken symmetry states in ABC-stacked trilayer graphene*, Phys. Rev. B **88**, 075408 (2013).

[51] Bocarsly, M. et al. *De Haas–van Alphen spectroscopy and magnetic breakdown in moiré graphene*, Science **383**, 42–48 (2024).

[52] Koshino, M. et al. *Maximally Localized Wannier Orbitals and the Extended Hubbard Model for Twisted Bilayer Graphene*. Phys. Rev. X **8**, 031087 (2018).

[53] Zhou, H. et al. *Half- and quarter-metals in rhombohedral trilayer graphene*, Nature **598**, 429–433 (2021).

[54] Z. Huo *et al.*, *Does Moire Matter? Critical Moire Dependence with Quantum Fluctuations in Graphene Based Integer and Fractional Chern Insulators*, arXiv:2510.15309.

## Method

### Device Fabrication

Atomic-thin flakes of graphite and hexagonal boron nitride (hBN) are obtained through mechanical exfoliation from bulk crystals. The tetralayer graphene with a rhombohedral stacking order is identified using near-field infrared microscopy and isolated from other stacking orders by AFM lithography (anodic oxidation). To improve the device yield, the rhombohedral tetralayer graphene was further cut into multiple pieces using AFM lithography. All stacks are assembled using a standard dry-transfer technique with a poly (bisphenol A carbonate) (PC)/polydimethylsiloxane (PDMS) stamp. To fabricate a controlled pair of devices with opposite hBN alignment orientations, we first prepared a large, uniform hBN flake and cut it into two pieces using a tungsten probe. We first picked up one hBN piece and used it to pick up one part of the graphene flake, intentionally aligning the straight edges with each other. The assembled stack is then released onto a prepared bottom stack, consisting of bottom hBN as the bottom dielectric and bottom graphite as the bottom gate on a $Si/SiO_2$ substrate. The bottom stack is also transferred using the dry transfer technique, followed by cleaning using annealing and AFM contact-mode. For the second stack, we picked up the remaining hBN piece with the same pick-up orientation as the first stack. The remaining graphene piece was then rotated 180° relative to its previous orientation before being picked up, forming the second stack. The surface of these two stacks was then cleaned by AFM contact mode, and top gates were subsequently released for both stacks, resulting in two devices with opposite hBN alignment orientations.

For other RMG/hBN devices, the hBN flakes were used without cutting, and the stacks were assembled following the standard dry transfer process. All stacks are shaped into a standard Hall bar geometry using e-beam lithography and reactive ion etching ($CHF_3/O_2$). Finally, electrical contacts are made by depositing Cr/Au (2/80 nm) metal edge contacts.

## Transport Measurement

Electrical transport measurements above 1.5 K were conducted in an Oxford Teslatron system. Devices 2-5 were further measured in a cryogen free dilution refrigerator (Q-one, Q-400) equipped with a 9 T superconducting magnet. Each fridge line has a sliver epoxy filter and multiple stage RC- filters at low temperature. The nominal base temperature is about 10 mK. Device 7 was measured in a top-loading dilution refrigerator (Oxford TLM) equipped with an 18 T superconducting magnet. The nominal base temperature is about 15 mK. Each fridge line has a sliver epoxy filter and a RC- filter at low temperature. A standard low-frequency lock-in technique with an excitation current of 1-3 nA at 3.77 Hz-23.37 Hz was used to measure longitudinal resistance $R_{xx}$ and Hall resistance $R_{xy}$.

To eliminate geometric mixing between longitudinal resistance $R_{xx}$ and Hall resistance $R_{xy}$, a standard symmetrization procedure was used to extract the symmetrized $R_{xx}$ and anti-symmetrized $R_{xy}$. Specifically, $R_{xx}(\pm B) = [R_{xx}(+B) + R_{xx}(-B)]/2$ and $R_{xy}(\pm B) = [R_{xy}(+B) - R_{xy}(-B)]/2$. We convert the measured $R_{xx}$ into longitudinal resistivity $\rho_{xx}$ by $\rho_{xx} = R_{xx}\frac{W}{L}$, where $W$ is the Hall bar width and $L$ is the separation between voltage probes.

## Calibration of moiré filling factors

The dual-gate geometry of the device allows us to independently tune the carrier density $\left(n = \frac{c_t V_t + c_b V_b}{e} + n_0\right)$ and the perpendicular displacement electric field $\left(D = \frac{c_t V_t - c_b V_b}{2\varepsilon_0} + D_0\right)$ in graphene layer by applying top graphite gate voltage $V_t$ and bottom graphite gate voltage $V_b$. Here, $\varepsilon_0$, $c_t$, $c_b$, $n_0$ and $D_0$ represents the vacuum permittivity, geometric capacitance of the top graphite gate, geometric capacitance of the bottom graphite gate, intrinsic doping and the built-in electric field, respectively. The value of $c_t$ and $c_b$ are mainly determined by measuring quantum oscillations. The RMG layer is aligned with one hBN layer to form a moiré superlattice. Based on the Brown-Zak oscillation measurements (see details in Extended Data Fig. 2), the moiré period $\lambda$ is determined, and the corresponding twist angle $\theta$ is extracted using $\lambda = \frac{(1+\delta)a}{\sqrt{2(1+\delta)[1-\cos(\theta)]+\delta^2}}$. Here, we take the lattice mismatch between hBN and graphene as $\delta \approx 1.7\%$ and the graphene lattice constant as $a \approx 0.246$ nm. The detailed information for all devices obtained through this procedure is summarized in Table I.

## Continuum model for rhombohedral tetralayer graphene aligned with hBN

The continuum model of RTG aligned with hBN is modeled by [47,48]

$$H^{(\xi)} = \begin{bmatrix} \sigma_x^{\xi} H_{\mathrm{hBN}} \sigma_x^{\xi} & T^{(\xi)}(r) \\ T^{(\xi)}(r)^{\dagger} & H_{\mathrm{RTG}} \end{bmatrix},$$

where $\xi$ denotes the orientation of the alignment, $H_{\mathrm{hBN}}$ and $H_{\mathrm{RTG}}$ respectively describe the intralayer Hamiltonian of hBN and RTG, and $T^{(\xi)}(r)$ gives the interlayer tunnels between the hBN and the top layer of the RTG, leading to a moiré superlattice.

Because the dispersion of hBN is negligible compared to the large potential of boron $V_{\mathrm{B}} = 3.4$ eV and nitrogen $V_{\mathrm{N}} = -1.4$ eV, the Hamiltonian is given by $H_{\mathrm{hBN}} = \mathrm{diag}(V_{\mathrm{B}}, V_{\mathrm{N}})$. The intralayer

Hamiltonian of RTG is composed of two parts, $H_{\mathrm{RTG}} = h_0 + V_{\mathrm{d}}$. The kinetic energy $h_0$ is given by [49,50],

$$h_0 = \begin{bmatrix} h^{(0)} & h^{(1)} & h^{(2)} & 0 \\ h^{(1)\dagger} & h^{(0)} & h^{(1)} & h^{(2)} \\ h^{(2)\dagger} & h^{(1)\dagger} & h^{(0)} & h^{(1)} \\ 0 & h^{(2)\dagger} & h^{(1)\dagger} & h^{(0)} \end{bmatrix},$$

where the Hamiltonian is written in the basis ($A_1$, $B_1$, $A_2$, $B_2$, $A_3$, $B_3$, $A_4$, $B_4$), and

$$h^{(0)} = \frac{3}{2} a_{\mathrm{G}} t_0 \begin{bmatrix} 0 & k^\dagger \\ k & 0 \end{bmatrix}, \qquad h^{(1)} = \begin{bmatrix} \frac{3}{2} a_{\mathrm{G}} t_4 k^\dagger & \frac{3}{2} a_{\mathrm{G}} t_3 k \\ t_1 & \frac{3}{2} a_{\mathrm{G}} t_4 k^\dagger \end{bmatrix}, \qquad h^{(0)} = \begin{bmatrix} 0 & t_2/2 \\ 0 & 0 \end{bmatrix},$$

with $k = k_x + ik_y$. In addition, the potential of each layer and sublattice is

$$V_{\mathrm{d}} = \Delta/2(-3,-3,-1,-1,1,1,3,3) + \delta_1(0,1,1,1,1,1,1,0) + \delta_2(-1,-1,1,1,1,1,-1-1),$$

where $\Delta$ is the potential difference between two adjacent layers caused by the displacement field, $\delta_1$ is the potential difference between the unpaired sublattices $A_1$, $B_4$ and the other sublattices, and $\delta_2$ is the potential difference between the outer layers and the other layers. For conduction bands, $\Delta < 0$ corresponds to the moiré proximal side, and vice versa. Following Ref. 40,51, we take

$$(t_0, t_1, t_2, t_3, t_4, \delta_1, \delta_2) = (3100, 380, -21, 290, 141, 10.5, 2)\ \mathrm{meV}.$$

The interlayer tunneling $T^{(\xi)}(r)$ is generated by the lattice mismatch and the twist angle between the graphene and the hBN. The reciprocal basis vectors of graphene are

$$G_1 = \frac{2\pi}{a_{\mathrm{G}}}[1, -1/\sqrt{3}], \qquad G_2 = \frac{2\pi}{a_{\mathrm{G}}}[0, 2/\sqrt{3}],$$

and the reciprocal basis vectors of the aligned hBN are

$$G_{\mathrm{i}}^{'} = \frac{a_{\mathrm{G}}}{a_{\mathrm{hBN}}} R_\theta G_{\mathrm{i}}$$

with a lattice mismatch $a_{\mathrm{G}}/a_{\mathrm{hBN}} \approx 1.018$. Here, $R_\phi$ is the counterclockwise rotation by angle $\phi$, and $\theta$ is the twist angle. Hence, the moiré reciprocal vectors are $g_i = G_i - G_i^{'}$. Similar to the twisted bilayer graphene, the tunneling can be modeled by [48, 51]

$$T^{(\xi)}(r) = T_0{}^{(\xi)} + T_1{}^{(\xi)} e^{ig_1 \cdot r} + T_2{}^{(\xi)} e^{ig_2 \cdot r}$$

with

$$T_j^{(\xi=1)} = \begin{bmatrix} \kappa t_{\mathrm{NC}} & t_{\mathrm{NC}} \omega^{-j} \\ t_{\mathrm{BC}} \omega^{\mathrm{j}} & \kappa t_{\mathrm{BC}} \end{bmatrix}, \ T_{\mathrm{j}}^{(\xi=0)} = \begin{bmatrix} \kappa t_{\mathrm{BC}} & t_{\mathrm{BC}} \omega^{-j} \\ t_{\mathrm{NC}} \omega^{j} & \kappa t_{\mathrm{NC}} \end{bmatrix},$$

where $t_{\mathrm{BC}}$ ($t_{\mathrm{NC}}$) is the tunneling between boron (nitrogen) atoms and carbon atoms, $\omega = e^{2\pi/3}$, and $\kappa$ results from lattice relaxation and corrugation [48, 51, 52].

As the energy scale of hBN is much higher than that of RTG, an effective model of RTG aligned with hBN can be obtained through the second-order perturbation theory,

$$H^{(\xi)} = H_{\mathrm{RTG}} + U^{(\xi)}(r) \tag{1}$$

with $U^{(\xi)}(r) = -T^{(\xi)}(r)^\dagger \left[\sigma_{\mathrm{x}}^{\xi} H_{\mathrm{hBN}} \sigma_{\mathrm{x}}^{\xi}\right]^{-1} T^{(\xi)}(r)$. The resulting moiré potential is explicitly given by

$$U^{(\xi=0)}(r) = U_{\mathrm{t}} + U_1 e^{ig_1\cdot r} + U_2 e^{ig_2\cdot r} + U_3 \mathrm{e}^{-i(g_1+g_2)\cdot r} + \text{H.c.},$$

where $U_t = \mathrm{diag}(V_{t1}, V_{t2})$ and

$$U_1 = \begin{bmatrix} V_1 & V_2/\omega \\ V_2 & V_3/\omega \end{bmatrix}, \qquad U_2 = \begin{bmatrix} V_1 & V_2\omega \\ V_2\omega & V_3/\omega \end{bmatrix}, \qquad U_3 = \begin{bmatrix} V_1 & V_2 \\ V_2/\omega & V_3/\omega \end{bmatrix}$$

with

$$V_{t1} = -\frac{3}{2}(\kappa^2 t_{\mathrm{BC}}^2/V_{\mathrm{B}} + t_{\mathrm{NC}}^2/V_{\mathrm{N}}), \qquad V_{t2} = -\frac{3}{2}(t_{\mathrm{BC}}^2/V_{\mathrm{B}} + \kappa^2 t_{\mathrm{NC}}^2/V_{\mathrm{N}})$$

$$V_1 = -\kappa^2 t_{\mathrm{BC}}^2/V_{\mathrm{B}} - \omega t_{\mathrm{NC}}^2/V_{\mathrm{N}},\ V_3 = -\kappa t_{\mathrm{BC}}^2/V_{\mathrm{B}} - \omega\kappa t_{\mathrm{NC}}^2/V_{\mathrm{N}},\ V_3 = -t_{\mathrm{BC}}^2/V_{\mathrm{B}} - \omega\kappa^2 t_{\mathrm{NC}}^2/V_{\mathrm{N}}$$

The moiré potentials for the two orientations are related through $U^{(\xi=0)}(r) = \sigma_{\mathrm{x}} U^{(\xi=1)}(-r)^* \sigma_{\mathrm{x}}$. We use the effective model Eq. (1) in all calculations. Following Ref. 40, 51, we take $t_{\mathrm{BC}} = 144$ meV, $t_{\mathrm{BC}} = 72$ meV and $\kappa = 0.5$.

The role of orientations is primarily represented by $|V_1|$ and $|V_3|$ in the presence of a strong displacement field. This is because the wave function in RTG is strongly polarized to $A_1$ and $B_4$ sublattices. Therefore, the moiré effect is dominantly controlled by the magnitude of the (1,1) entry of $U^{(\xi)}(r)$, which is $|V_1|$ for $\xi = 0$ and $|V_3|$ for $\xi = 1$. We further note that $U_t$ does not play a significant role, as it is a constant potential. Without relaxation, we have $\kappa = 1$, so $|V_1| = |V_3|$ and the spectrum is almost identical for the two orientations. Thus, relaxation is critical for the discrepancy between the two orientations and results in two disparate spectra in Extended Data Fig. 7. Particularly, the $\xi = 1$ orientation has a sizable global gap above the lowest conduction band, while the $\xi = 0$ orientation does not.

**Hartree-Fock theory**

For the Hartree-Fock calculation, we consider the following interaction,

$$V_{\mathrm{int}} = \frac{1}{2A}\int \mathrm{d}q\, V(q) : \rho(q)\rho(-q) : -J\int \mathrm{d}r : \vec{\boldsymbol{\sigma}}_{\mathrm{K}}(r)\cdot\vec{\boldsymbol{\sigma}}_{\mathrm{K}'}(r') :$$

where $A$ is the area of the system, the first term is the Coulomb potential, and the second term is the Hund's term [53]. Here, $\rho(q)$ is the Fourier transform of the total density of the four flavors, and $\vec{\sigma}_K(r)$, $\vec{\sigma}_{K'}(r')$ are the local spin operator of each valley. The gate-screened Coulomb potential $V(q)$ is given by

$$V(q) = \frac{2\pi e^2 \tanh qd}{\varepsilon q},$$

where $d = 30$ nm is the distance between the two gates, and the dielectric constant is taken to be $\varepsilon = 10$. Further, we take $J/A_{\mathrm{moiré}} = 2$ meV, where $A_{\mathrm{moiré}}$ is the area of the moiré unit cell. This corresponds to an energy decrease of 2 meV per electron when there is one electron per moiré unit cell in both $K\uparrow$ and $K'\downarrow$. Hence, $J > 0$ favors a spin-polarized state. We also note that $J$ used here is comparable to the Hund's coupling used in Ref. 53.

In Extended Data Fig. 7, we project the system onto the lowest five single-particle conduction bands for each flavor and perform the HF calculation at integer fillings on an 18×18 momentum mesh. The projection is valid for a strong displacement field, which creates a sizable band gap between the conduction and valence bands. Here, we take $\Delta = -55$ meV and $\theta = 0.48°$. For $\xi =$ 1, the interaction further increases the single-particle band gap. The ground state is a spin valley-polarized insulator at $\nu = 1$ and 3, a spin-polarized and valley-balance insulator at $\nu = 2$, and a band insulator at $\nu = 4$. We also note that the energy splitting of the three unoccupied flavors at $\nu = 1$ and the three occupied flavors at $\nu = 3$ is about 2 meV, stemming from the Hund's term. For $\xi = 0$, though the interaction can open a correlated gap at $\nu = 1$ and 3, the spin-polarized half metal is more energetically favored at $\nu = 3$ and 4. Furthermore, the correlated gap at $\nu = 1$ and 2 for $\xi = 0$ is much smaller than that for $\xi = 1$, in agreement with the experimental results.

**Note added**—During the preparation of this manuscript, we became aware of another complementary and independent study based on rhombohedral hexalayer graphene/hBN moiré system [54].

**Acknowledgement**

We acknowledge Prof. E. Zeldov, Prof. X. Lu, and Prof. N. Regnault for helpful discussions. This work is supported by the National Key R&D Program of China (Nos. 2022YFA1402404, 2022YFA1402702, 2022YFA1405400, 2021YFA1401400, 2021YFA1400100, 2020YFA0309000, 2019YFA0308600, 2022YFA1402401, 2020YFA0309000), the National Natural Science Foundation of China (Nos. 12350403, 12374045, 12174249, 12174250, 12141404, 12350005, 12174248, 92265102), the Innovation Program for Quantum Science and Technology (Nos. 2021ZD0302600 and 2021ZD0302500), Natural Science Foundation of Shanghai (No. 24QA2703700), Shanghai Municipal Science and Technology Major Project (grant No.2019SHZDZX01) and Shanghai Science and Technology Innovation Action Plan (grant No. 24LZ1401100). C.L., K.L., and F.X. are supported by T.D. Lee scholarship. X. Liu, T.L. acknowledge the Shanghai Jiao Tong University 2030 Initiative Program. X. Liu acknowledges "Shuguang Program" supported by Shanghai Education Development Foundation and Shanghai Municipal Education Commission. T.L. and G.C. acknowledges the Yangyang Development Fund. X. Li acknowledges the support from the Research Grants Council of Hong Kong (Grants No. 11312825, No. C7012-21G, and No. C7015-24G) and City University of Hong Kong (Project No. 9610428). K. W. and T. T. acknowledge support from the JSPS KAKENHI (Nos. 21H05233 and 23H02052) and World Premier International Research Center Initiative (WPI), MEXT, Japan. A portion of this work was performed at the Synergetic Extreme Condition User Facility (SECUF).

**Competing interests**

The authors declare no competing financial interests.

**Data availability**

All data that support the findings of this study are available from the contact author upon request.

# Figures

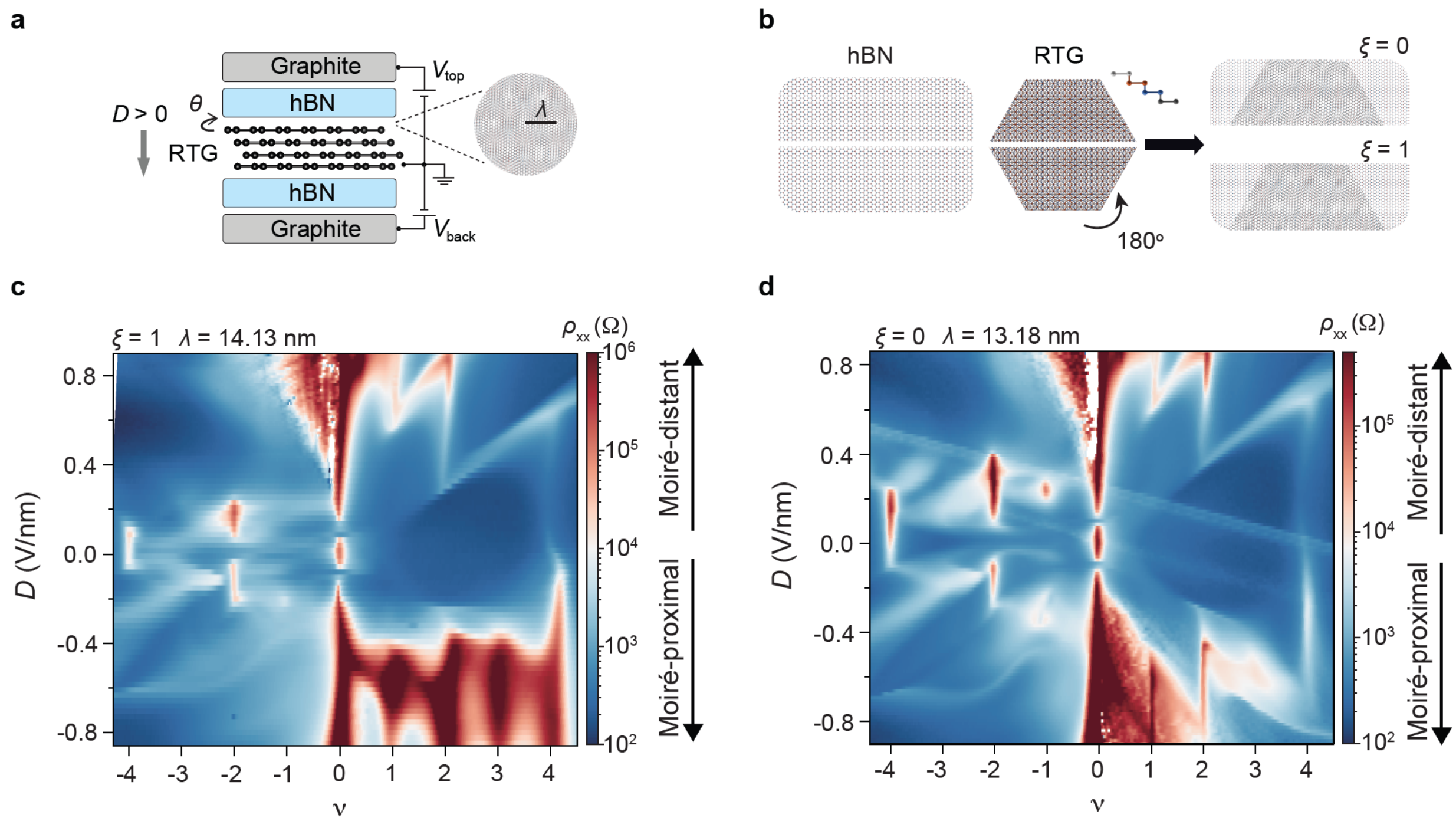

**Fig. 1. Device structure and comparison of transport behaviors between devices with opposite hBN alignment orientations. a**, Schematic of the RTG/hBN moiré device with dual graphite gates. The moiré superlattice is formed between the graphene layer and the top hBN, with the moiré period labeled as $\lambda$. **b**, Schematic of the fabrication process for a controlled pair of devices with opposite hBN alignment orientations. Uniform hBN and RTG flakes were cut into two pieces, and the first hBN and RTG flakes were directly assembled to form a moiré superlattice, while for the second device, the RTG flake was rotated by 180° before being picked up onto the hBN, resulting in an opposite alignment orientation. See Methods and Extended Data Fig. 1 for details of the device fabrication. **c**, **d**, Longitudinal resistivity $\rho_{xx}$ as a function of displacement electric field $D$ and moiré filling factor $\nu$ for RTG/hBN devices with alignment orientations $\xi = 1$ (moiré period $\lambda = 14.13$ nm, device 1, **c**) and $\xi = 0$ ($\lambda = 13.18$ nm, device 2, **d**), measured at $B = 0$ T and $T = 1.7$ K. The moiré-proximal and moiré-distant regions for electrons correspond to the negative-$D$ and positive-$D$ regions, respectively.

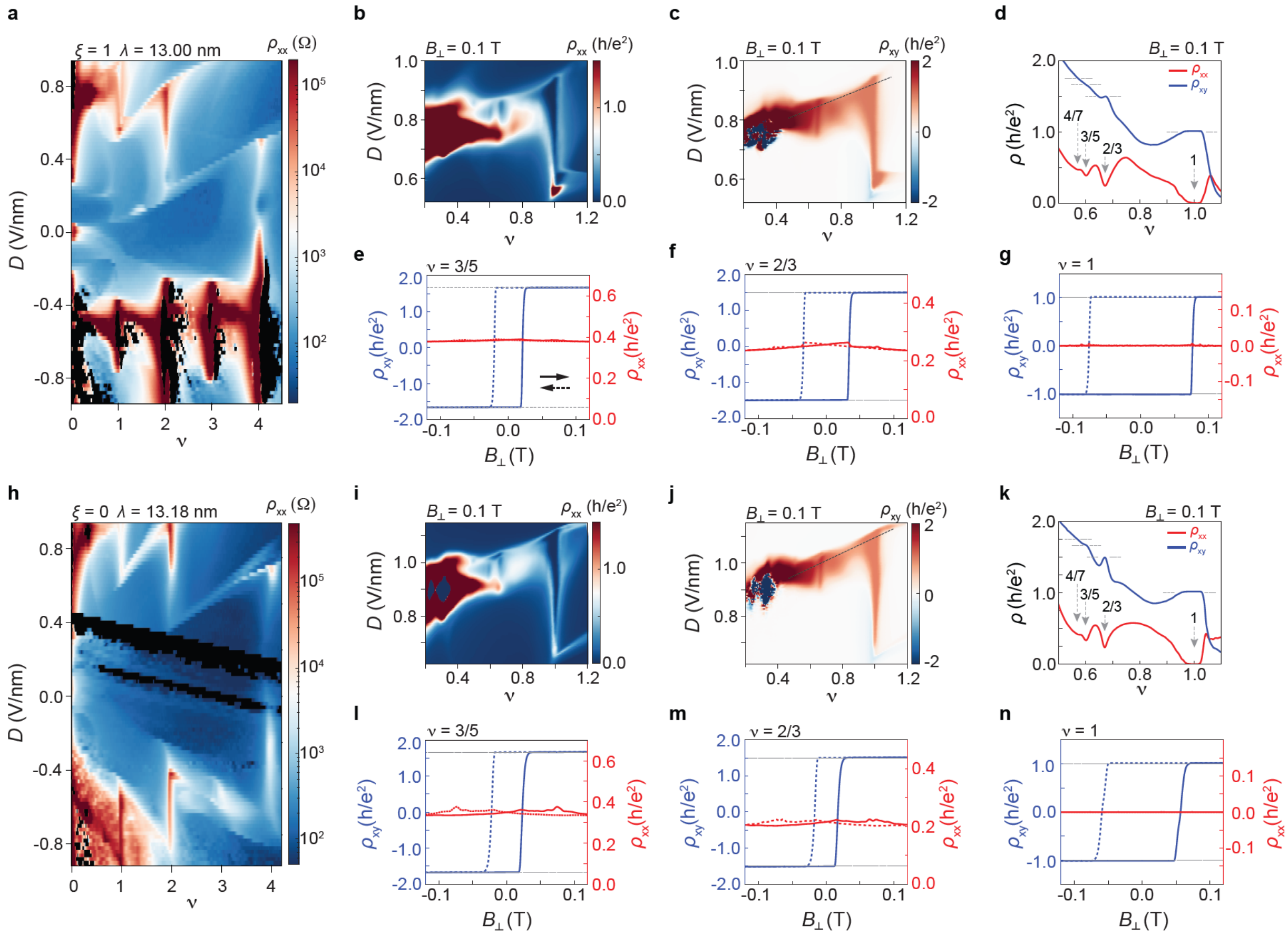

**Fig. 2. The IQAH and FQAH effects in the moiré-distant region of RTG/hBN with $\xi = 1$ and $\xi = 0$ alignment orientations. a**, Longitudinal resistivity $\rho_{xx}$ as a function of $D$ and $\nu$ in an RTG/hBN device with alignment orientation $\xi = 1$ and a moiré period of 13.0 nm (device 3), measured at $B_\perp = 0.1$ T and $T = 10$ mK. **b**,**c**, Symmetrized $\rho_{xx}$ (**b**) and anti-symmetrized $\rho_{xy}$ (**c**) measured at $B_\perp = \pm 0.1$ T and $T = 10$ mK, as a function of $\nu$ and $D$ in the moiré-distant region of device 3. **d**, Linecuts of $\rho_{xx}$ and $\rho_{xy}$ as a function of $\nu$ along the dashed line in **c**, showing the IQAH state at $\nu = 1$, and FQAH states at $\nu = 2/3$, 3/5 and 4/7. **e-g**, Magnetic hysteresis loops of $\rho_{xx}$ and $\rho_{xy}$ at $\nu = 3/5$ (**e**), $\nu = 2/3$ (**f**) and $\nu = 1$ (**g**), measured at $T = 10$ mK for device 3, with $\rho_{xy}$ exhibiting the quantized value of $h/\nu e^2$, as indicated by the grey dashed lines. **h**, $\rho_{xx}$ as a function of $D$ and $\nu$ in an RTG/hBN device with alignment orientation $\xi = 0$ and a moiré period of 13.18 nm (device 2), measured at $T = 300$ mK and $B_\perp = 0$ T. **i-n**, Similar plots for device 2, presenting the same measurements as shown in **b-g**, respectively. The black regions in **a**, **h** are highly insulating areas or regions with poor contacts.

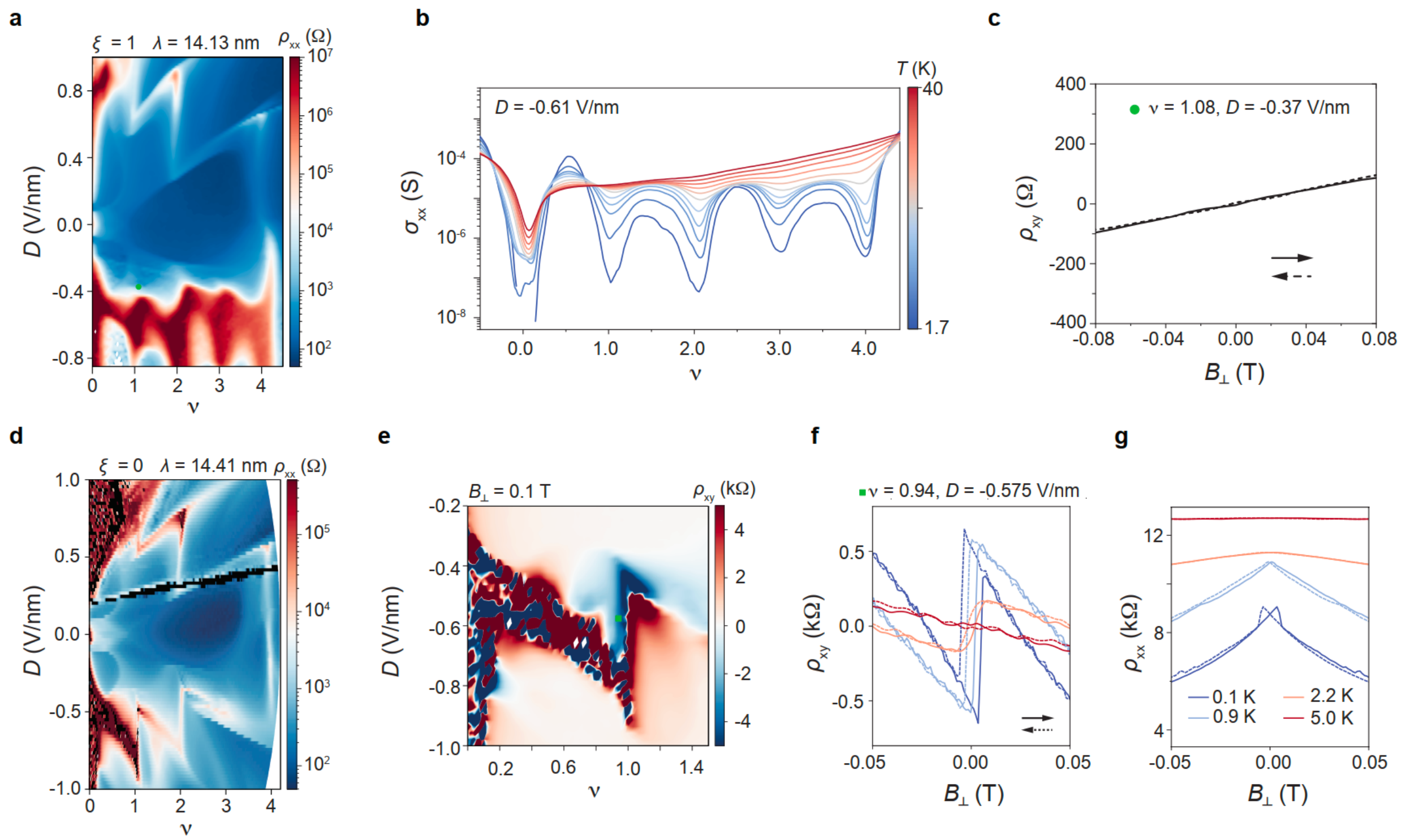

**Fig. 3. Distinct transport properties in the moiré-proximal region for opposite hBN alignment orientations. a,** $\rho_{xx}$ as a function of $D$ and $\nu$ in an RTG/hBN device with alignment orientation $\xi = 1$ and a moiré period of 14.13 nm (device 1), measured at $T = 300$ mK and $B_\perp = 0$ T. **b,** Temperature dependence of $\sigma_{xx}$ as a function of $\nu$ at fixed $D = -0.61$ V/nm, where all integer fillings ($\nu = 1, 2, 3, 4$) exhibit insulating behavior. **c**, Representative $\rho_{xy}$ as a function of $B_\perp$ at $\nu = 1.08$ and $D = -0.37$ V/nm (marked in **a**), measured at $T = 1.7$ K in the device shown in **a**. No magnetic hysteresis loop is observed, indicating the absence of orbital ferromagnetism. **d,** $\rho_{xx}$ as a function of $D$ and $\nu$ in an RTG/hBN device with alignment orientation $\xi = 0$ and a moiré period of 14.41 nm (device 4), measured at $T = 10$ mK and $B_\perp = 0.1$ T. **e**, Anti-symmetrized $\rho_{xy}$ as a function of $D$ and $\nu$ measured at $B_\perp = \pm 0.1$ T. **f,g**, Magnetic hysteresis loops of $\rho_{xy}$ (**f**) and $\rho_{xx}$ (**g**) at $\nu = 0.94$ and $D = -0.575$V/nm (marked in **e**), measured at various temperatures. The black regions in **d** are highly insulating areas or regions with poor contacts.

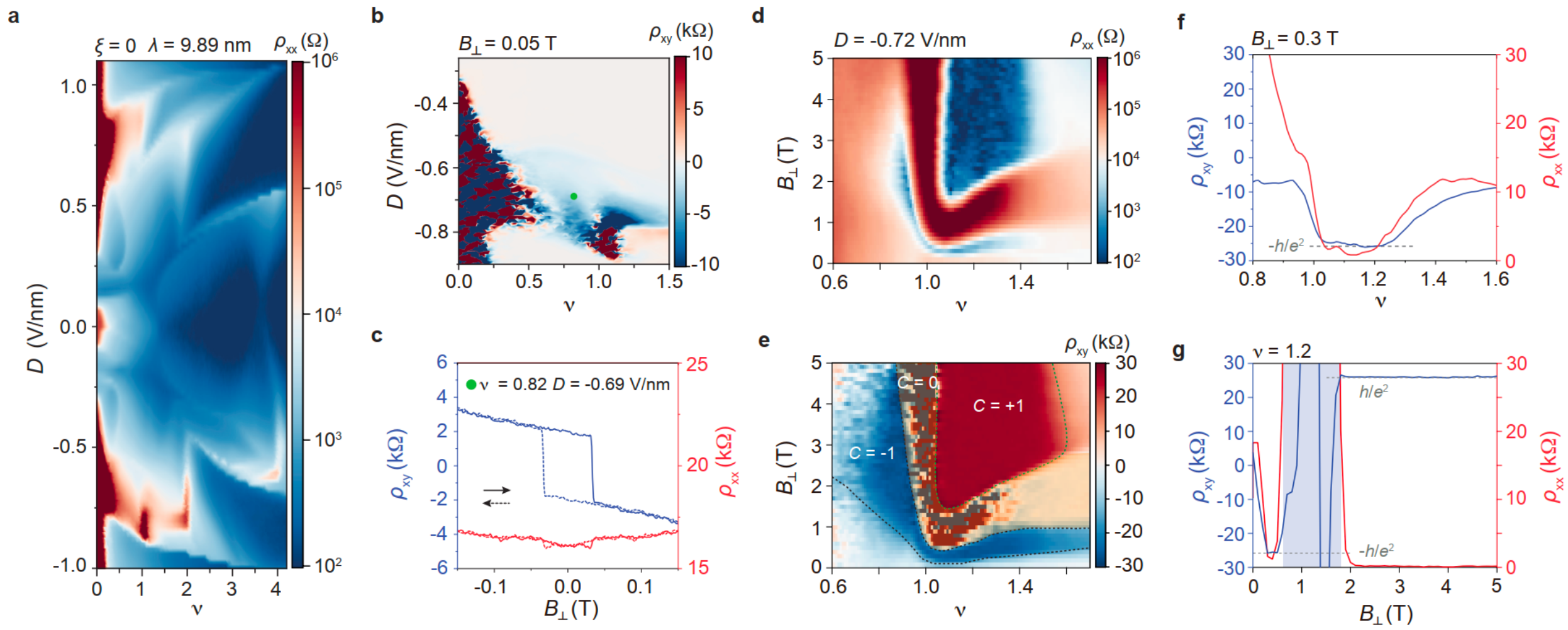

**Fig. 4. Chern insulators and trivial correlated insulators in the moiré-proximal region for $\xi$ = 0 hBN alignment**. **a,** $\rho_{xx}$ as a function of $D$ and $\nu$ in an RPG/hBN device with alignment orientation $\xi = 0$ and a moiré period of 9.89 nm (device 7), measured at $T = 20$ mK and $B_\perp = 0$ T. **b**, Anti-symmetrized $\rho_{xy}$ as functions of $D$ and $\nu$ at $B_\perp = \pm 0.05$ T. **c**, Representative magnetic hysteresis loops of $\rho_{xx}$ and $\rho_{xy}$ at $\nu = 0.82$ and $D = -0.69$ V/nm (marked in **b**), measured at $T = 20$ mK. **d, e,** Symmetrized $\rho_{xx}$ (**d**) and anti-symmetrized $\rho_{xy}$ (**e**) maps as functions of $B_\perp$ and $\nu$ measured at a fixed $D = -0.72$ V/nm. Integer Chern insulators with Chern numbers $C = -1$ (marked by the black dashed line) and $C = +1$ (marked by the green dashed line) emerge near $\nu = 1$ in the moiré-proximal region. The region highlighted by a transparent pattern corresponds to a trivial insulating state with $C = 0$. **f**,**g**, Line cuts of $\rho_{xx}$ and $\rho_{xy}$ as a function of $\nu$ at $B_\perp = 0.3$ T (**f**), and as a function of $B_\perp$ at $\nu = 1.2$ (**g**). The trivial correlated insulator region is marked by light blue, where both the $\rho_{xx}$ and $\rho_{xy}$ are highly resistive.

**Table I. Summary of device characteristics**

| Device | hBN alignment | Layer | Twist angle (degree) | Moiré wavelength (nm) | The $D$ range of the IQAH effect at $v$ =1 in the moiré-distant region (V/nm) | FCI states in the moiré-distant region | The AHE near $v$ =1 in the moiré-proximal region |
|---|---|---|---|---|---|---|---|
| D1 | $\xi = 1$ | 4 | 0.28 | 14.13 | 0.69-1.05 | Yes | No |
| D2 | $\xi = 0$ | 4 | 0.48 | 13.18 | 0.715-1.14 | Yes | Poor contact |
| D3 | $\xi = 1$ | 4 | 0.52 | 13.00 | 0.57-0.95 | Yes | No |
| D4 | $\xi = 0$ | 4 | 0.20 | 14.41 | Poor contact | Poor contact | Yes |
| D5 | $\xi = 1$ | 4 | 0.66 | 12.15 | 0.685-0.985 | Yes | No |
| D6 | $\xi = 1$ | 4 | 0.78 | 11.45 | 0.925-1.13 | No | No |
| D7 | $\xi = 0$ | 5 | 1.06 | 9.89 | Not quantized at $B$=0 | No | Yes |
| D8 | $\xi = 0$ | 5 | 0.09 | 14.66 | 0.64~0.85 | Poor contact | Yes |

Note: (i) The $D$ range of the IQAH effect at $v = 1$ is nominally defined as the range where the Hall resistance at $B_{\perp} = 0.1$ T reaches 80% of $h/e^2$. (ii) For extracting this $D$ range, all devices except D8 are based on data measured at $T = 10$ mK, while D8 is determined from data measured at $T = 1.7$ K.

## Extended Data Figures

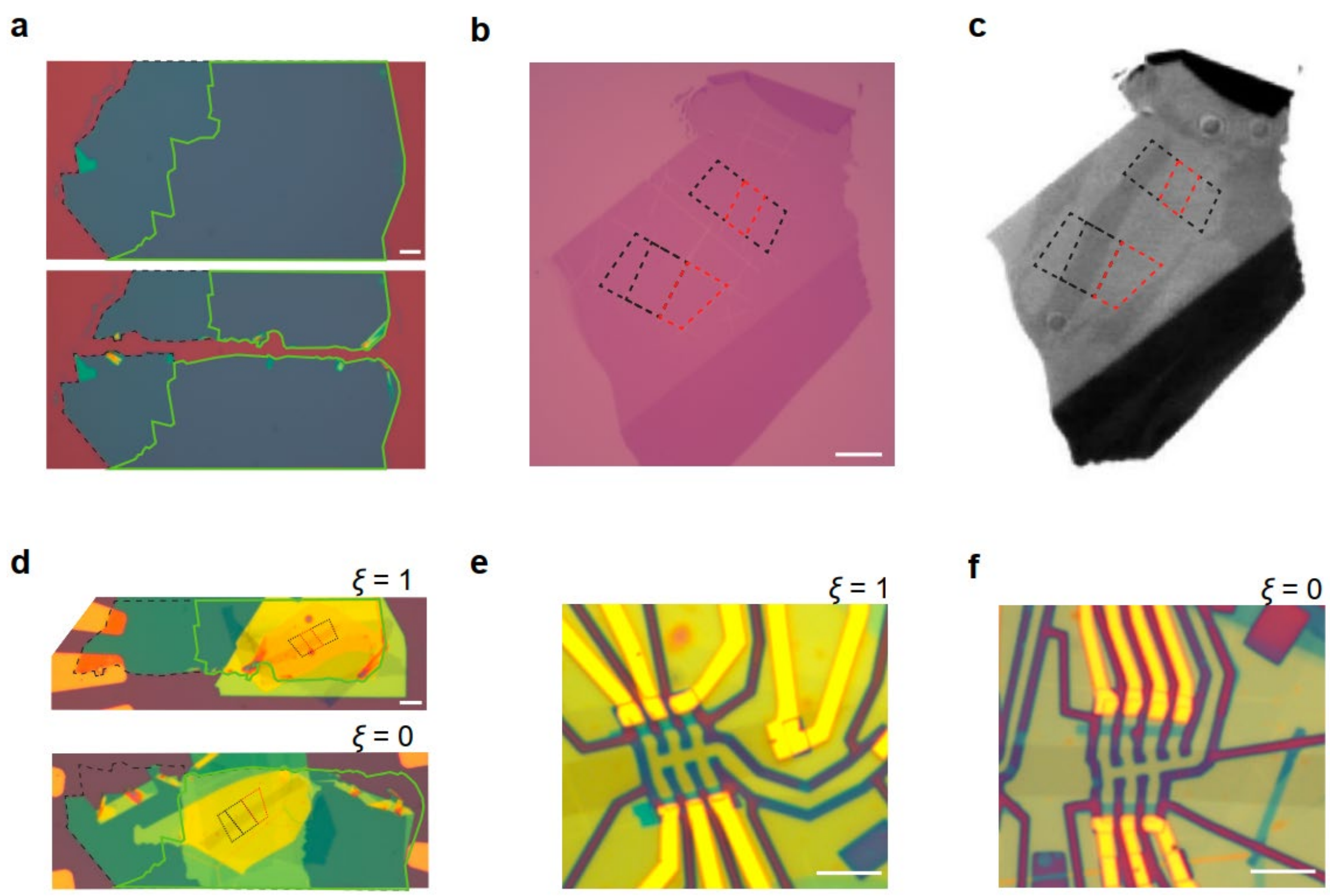

**Extended Data Fig. 1**. **Device fabrication procedures for device 1 and device 2. a**, Optical microscope images of the top hBN flake before (upper panel) and after (lower panel) cutting with a tungsten probe. The uniform thickness region is highlighted by green lines. **b**, **c,** Optical (b) and infrared (c) microscope images of a tetralayer graphene flake. The rhombohedral-stacked region is bright, while the Bernal-stacked region is dark in **c**. The rhombohedral-stacked region is isolated from the Bernal-stacked region by AFM lithography; the cutting path is marked by a dashed line, and the rhombohedral-stacked graphene pieces used in our devices are highlighted in red. These two graphene pieces, taken from the same flake, are picked up in two separate steps. Specifically, the two hBN pieces are picked up with the same orientation, while the second graphene piece is rotated by 180° relative to the first one before pickup. **d**, Optical microscope images of the stacks for the $\xi = 1$ and $\xi = 0$ devices. **e, f,** Optical microscope images of device 1 (**e**) and device 2 (**f**). They are shaped into Hall-bar geometries. Scale bars: 10 μm for **a-d** and 5 μm for **e-f**.

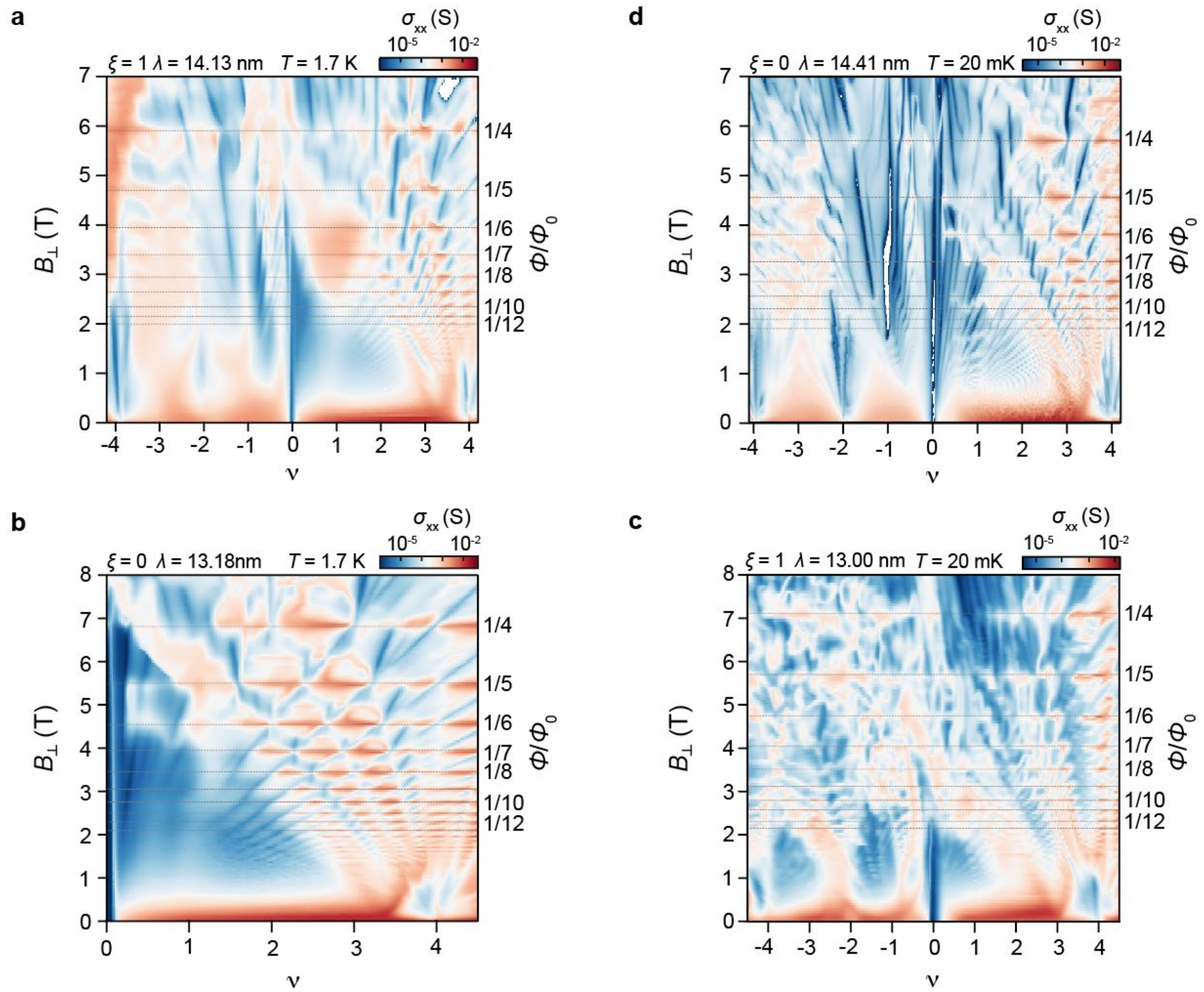

**Extended Data Fig. 2 Moiré periodicities determined from the Brown-Zak oscillations. a-d,** Conductivity $\sigma_{xx}$ as a function of moiré filling factors $\nu$ and perpendicular magnetic field $B_\perp$ for devices 1-4, respectively, with the corresponding hBN alignment orientations and moiré periods $\lambda$ labeled in the figure, where $\sigma_{xx} = \rho_{xx}/(\rho_{xx}^2 + \rho_{xy}^2)$. Brown–Zak oscillations appear when the magnetic flux through a moiré unit cell, $\Phi$, is a rational fraction of the magnetic flux quanta, i.e., $\Phi/\Phi_0 = p/q$, where $p$ and $q$ are integers. In each panel, the values of $\Phi/\Phi_0$ are labeled on the right axis. Fitting these $\Phi/\Phi_0$ values provides the area of the moiré unit cell $S$, and the full-filling density of the first moiré band is then given by $n_m = 4/S$. The moiré period λ is further determined from $n_m = \frac{8}{\sqrt{3}\lambda^2}$.

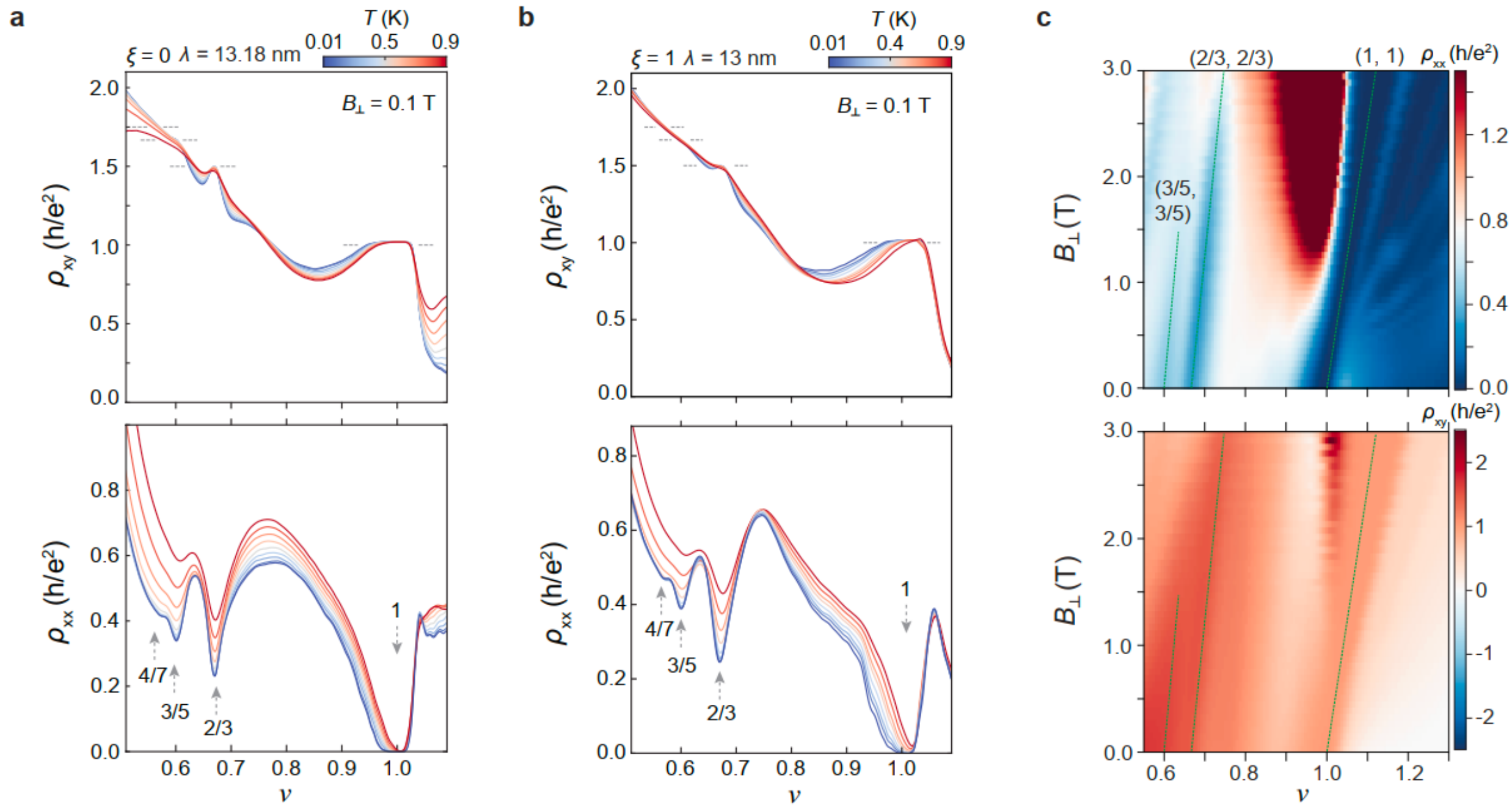

**Extended Data Fig. 3 Additional characterizations of the IQAH and FQAH effects in the moiré-distant region of device 2 and device 3. a, b,** Temperature dependence of $\rho_{xx}$ and $\rho_{xy}$ as a function of moiré fillings along the dashed lines shown in Fig. 2c (device 3) and 2j (device 2), respectively. The IQAH and FQAH states are indicated in the figures. **c**, Landau fan diagram measured in device 3, measured at $D$ = 0.855 V/nm and $T$ = 10 mK. The IQAH state at $\nu$ = 1 and the FQAH states at $\nu$ = 3/5 and 2/3 are indicated by ($\nu$, $C$), where $\nu$ is the moiré filling factor and $C$ is the Chern number; all states follow the Streda formula $n_m \frac{d\nu}{dB} = C\frac{e}{h}$, as indicated by green dashed lines in **c**. Notably, the FQAH states are gradually suppressed by the applied $B_\perp$: the $\nu$ = 3/5 state disappears at $B_\perp > 1.5$ T, while the $\nu$ = 2/3 state disappears at $B_\perp > 3$ T.

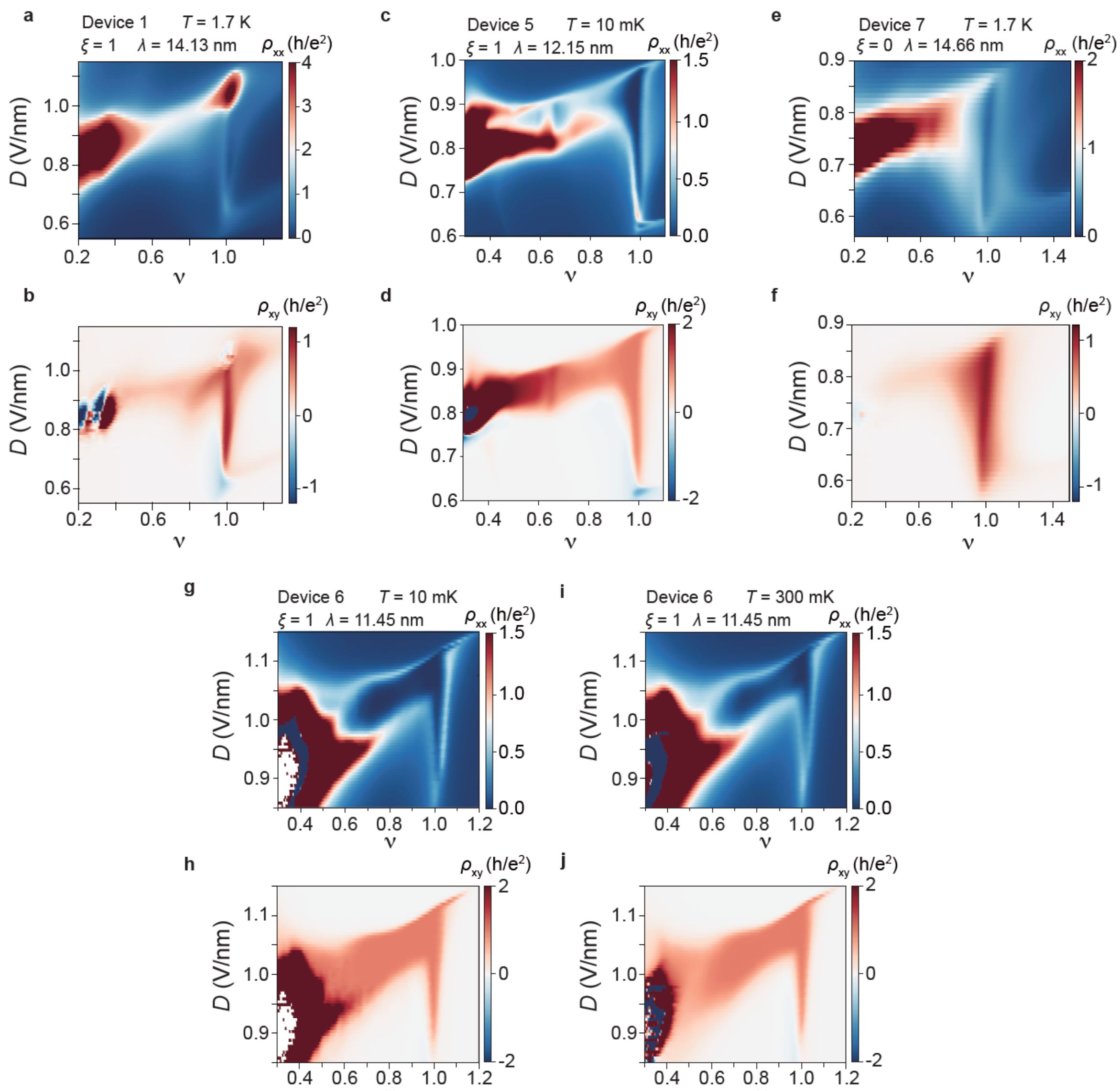

**Extended Data Fig. 4. Additional $\nu$-D maps in the moiré-distant region from other devices.** **a, b,** Symmetrized $\rho_{xx}$ (**a**) and antisymmetrized $\rho_{xy}$ (**b**) as functions of moiré filling factor $\nu$ and displacement electric field $D$ measured in device 1 at $T = 1.7$ K. **c, d,** $\rho_{xx}$ (**c**) and $\rho_{xy}$ (**d**) as a function of $\nu$ and $D$ measured in device 5 at $T = 10$ mK. **e, f,** Symmetrized $\rho_{xx}$ (**e**) and antisymmetrized $\rho_{xy}$ (**f**) as a function of ν and $D$ measured in device 8 at $T = 1.7$ K. **g, h,** $\rho_{xx}$ (**g**) and $\rho_{xy}$ (**h**) as a function of $\nu$ and $D$ measured in device 6 at $T = 10$ mK. **i, j,** Symmetrized $\rho_{xx}$ (**i**) and antisymmetrized $\rho_{xy}$ (**j**) as a function of ν and $D$ measured in device 6 at $T = 300$ mK. Note that, panels **g-h** are measured with an ac excitation current of 1 nA, while panels **i-j** are measured with 3 nA. All maps are measured at $B_{\perp} = 0.1$ T. The IQAH effect at $\nu = 1$ is observed in all these devices. In device 6, $\rho_{xy}$ shows a quantized value of $h/e^2$ accompanied by a vanishing $\rho_{xx}$ over a wide range of moiré filling factors, and no FQAH states are observed even with increasing temperature or excitation current. This behavior differs from the previously reported extended quantum anomalous Hall effect [12]. The origin of this difference may be associated with the twist-angle-dependent band structures and many-body effects, which require further investigation.

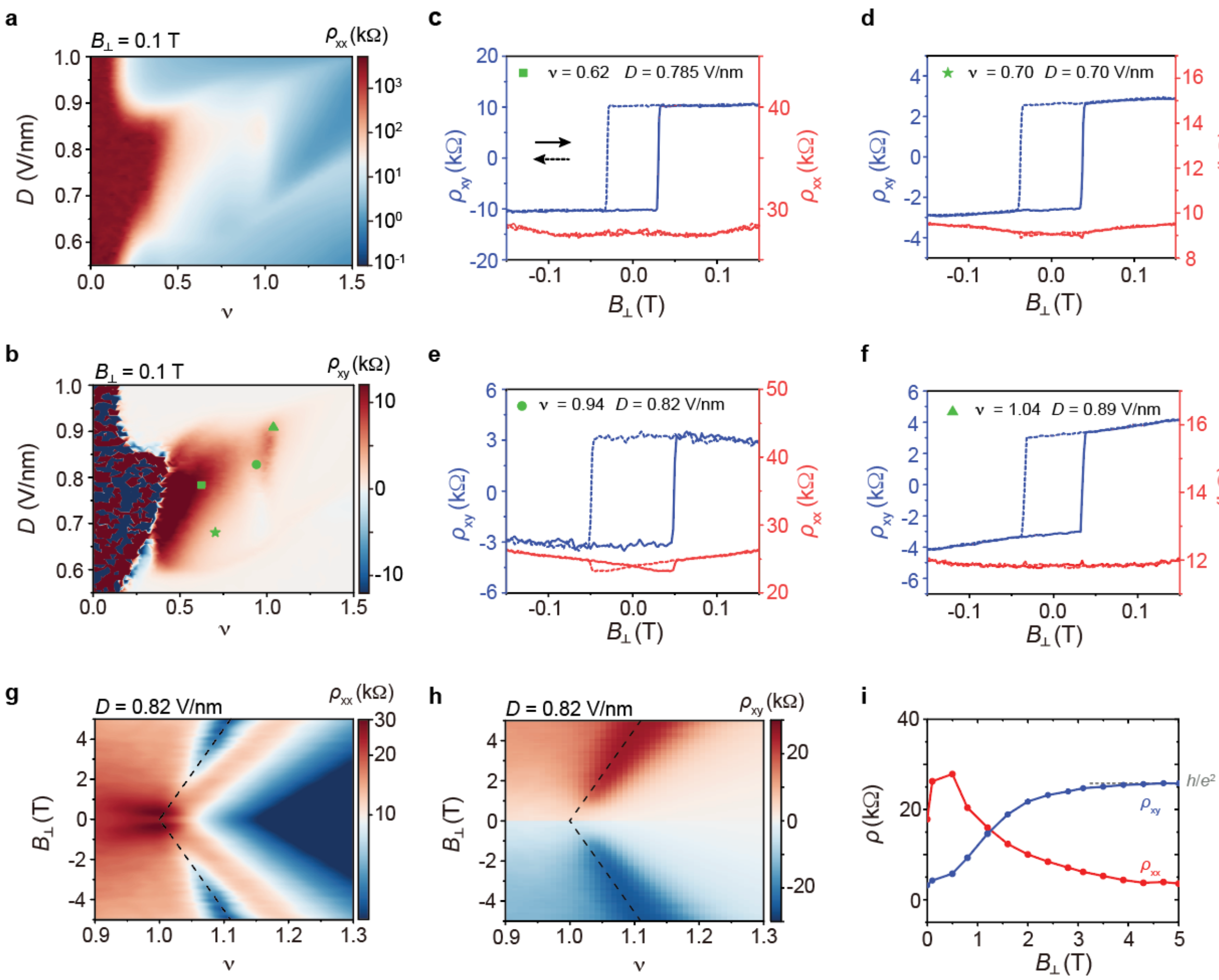

**Extended Data Fig. 5 The AH effect in the moiré-distant region of device 7 (RPG/hBN, $\lambda$ = 9.89 nm, $\xi$ = 0). a, b,** Maps of the symmetrized $\rho_{xx}$ (**a**) and anti-symmetrized $\rho_{xy}$ (**b**) as functions of $D$ and $\nu$, measured at $B_\perp = \pm 0.1$ T in device 7. **c-f** Magnetic hysteresis loops of $\rho_{xx}$ and $\rho_{xy}$ measured at different $\nu$ and $D$, as labeled in the **b**, respectively. **g**, **h**, Symmetrized $\rho_{xx}$ (**g**) and anti-symmetrized $\rho_{xy}$ (**h**) as functions of $B_\perp$ and $\nu$ measured at $D$ = 0.82 V/nm. The ICI state, characterized by vanishing in $\rho_{xx}$ and quantized $\rho_{xy} = h/e^2$, is stabilized by applied $B_\perp$. Its evolution in the $\nu$ - $B$ map follows the Streda formula, as indicated by the black dashed line. **i**, Line cuts of $\rho_{xx}$ and $\rho_{xy}$ along the dashed line in **g**–**h**.

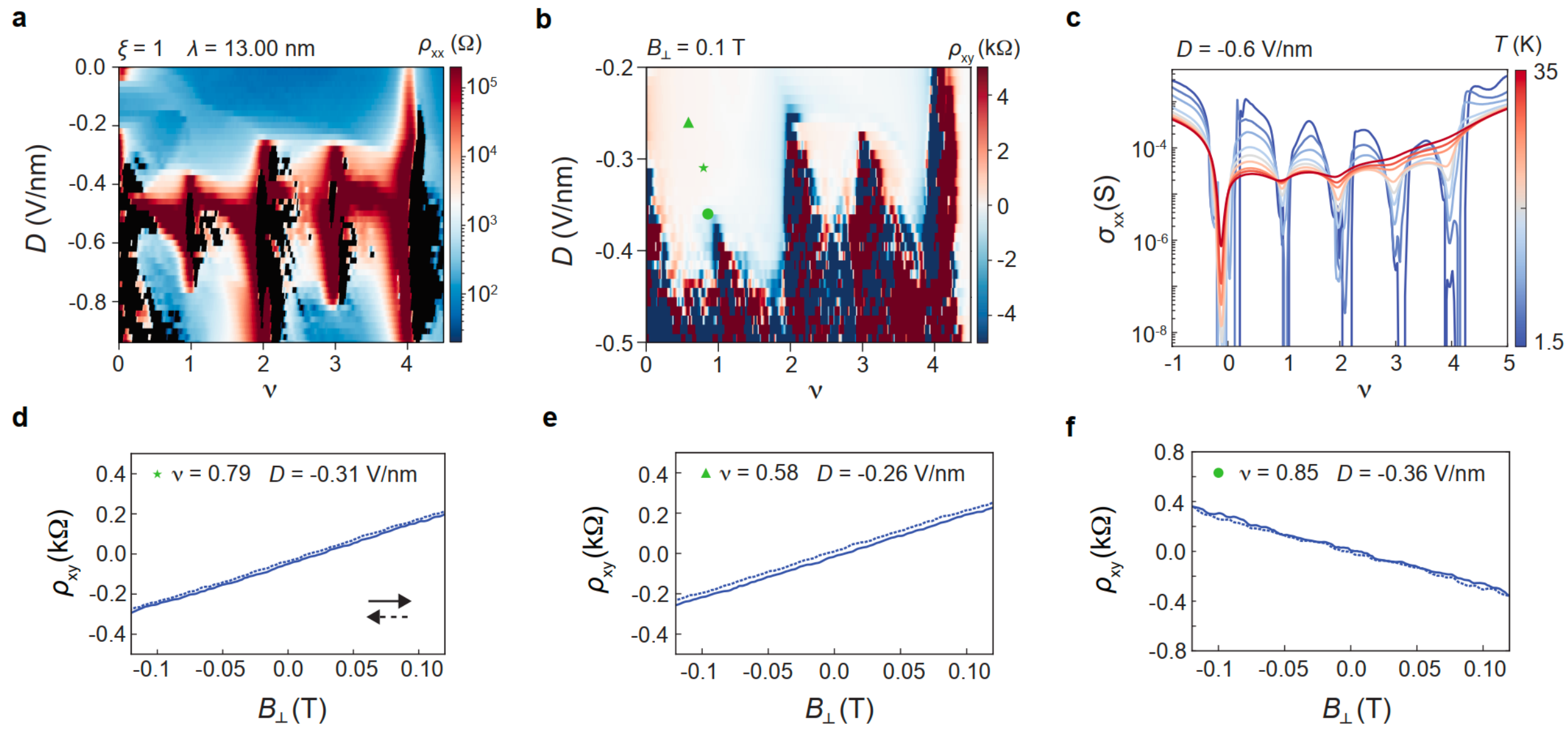

**Extended Data Fig. 6 Additional characterizations of device 3 ($\xi = 1$ alignment) in the moiré-proximal region. a, b,** $\rho_{xx}$ (**a**) and anti-symmetrized $\rho_{xy}$ (**b**) as a function of $D$ and $\nu$ in device 3, measured at $B_\perp = 0.1$ T and $T = 20$ mK. c, Temperature dependence of the conductivity $\sigma_{xx}$ as a function of $\nu$, measured in the same device as with fixing $D$ = -0.6 V/nm. Obviously, strong insulating behavior is observed at all integer moiré filling factors. **d-f**, Representative magnetic hysteresis scans measured at $T = 20$ mK, with the corresponding ($\nu$, $D$) points marked in **b**. No magnetic hysteresis loops are observed in **d-f**.

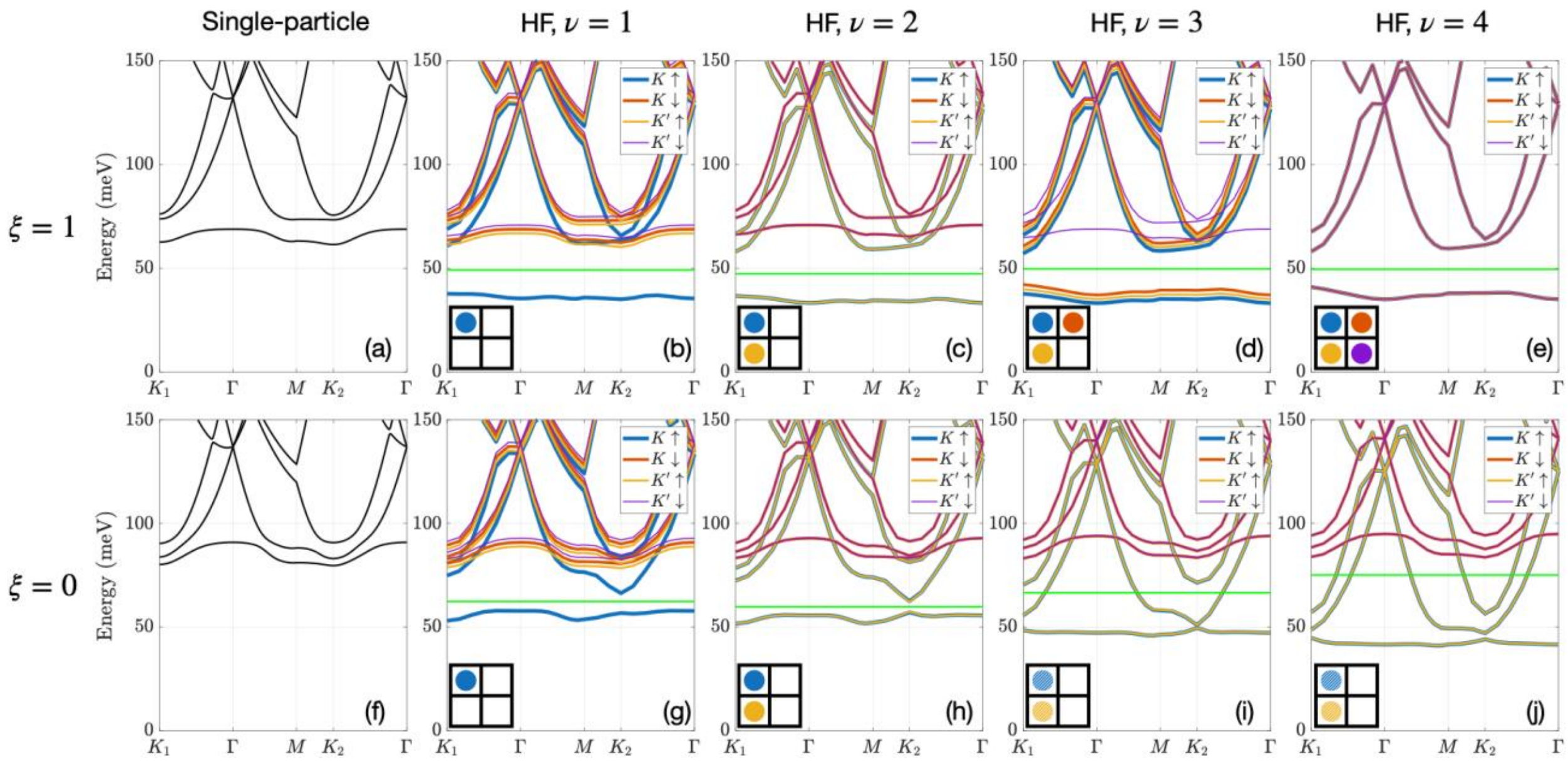

**Extended Data Fig. 7. Band Structure Calculations. a (f),** Theoretical calculations of single-particle band structures for RTG/hBN with $\xi$ = 1 ($\xi$ = 0) alignment, where the lowest moiré conduction band is isolated only in the $\xi$ = 1 alignment case. **b-e (g-j),** Calculated HF band structures at $\nu$ = 1, 2, 3, and 4 for $\xi$ = 1 ($\xi$ = 0) alignment. The color represents the four flavors, and the green horizontal lines delineate the Fermi level. The bands in the *K* valley are plotted along the path ($K_1$, Γ, *M*, $K_2$, Γ) and the bands in the *K'* valley are plotted along the path (-$K_1$, Γ, -*M*, -$K_2$, Γ), so that the spectra of the four flavors exactly overlap in the absence of spontaneous time-reversal symmetry breaking. In the HF calculations, we retain five single-particle conduction bands for each flavor and perform the calculations on an 18 by 18 momentum mesh in the moiré Brillouin zone. Here, we take twist angle $\theta$ = 0.48°, Δ = -55 meV, dielectric constant $\varepsilon$ = 10, and Hund's coupling $J/A_{moiré}$=2 meV, where $A_{moiré}$ is the area of the moiré unit cell. In the insets of **b**–**e** and **g**–**h**, each solid-colored circle denotes complete filling of the lowest quasiparticle band for the indicated flavor; in panels **i**,**j**, shaded dots indicate that those two flavors are instead filled into higher quasiparticle bands.

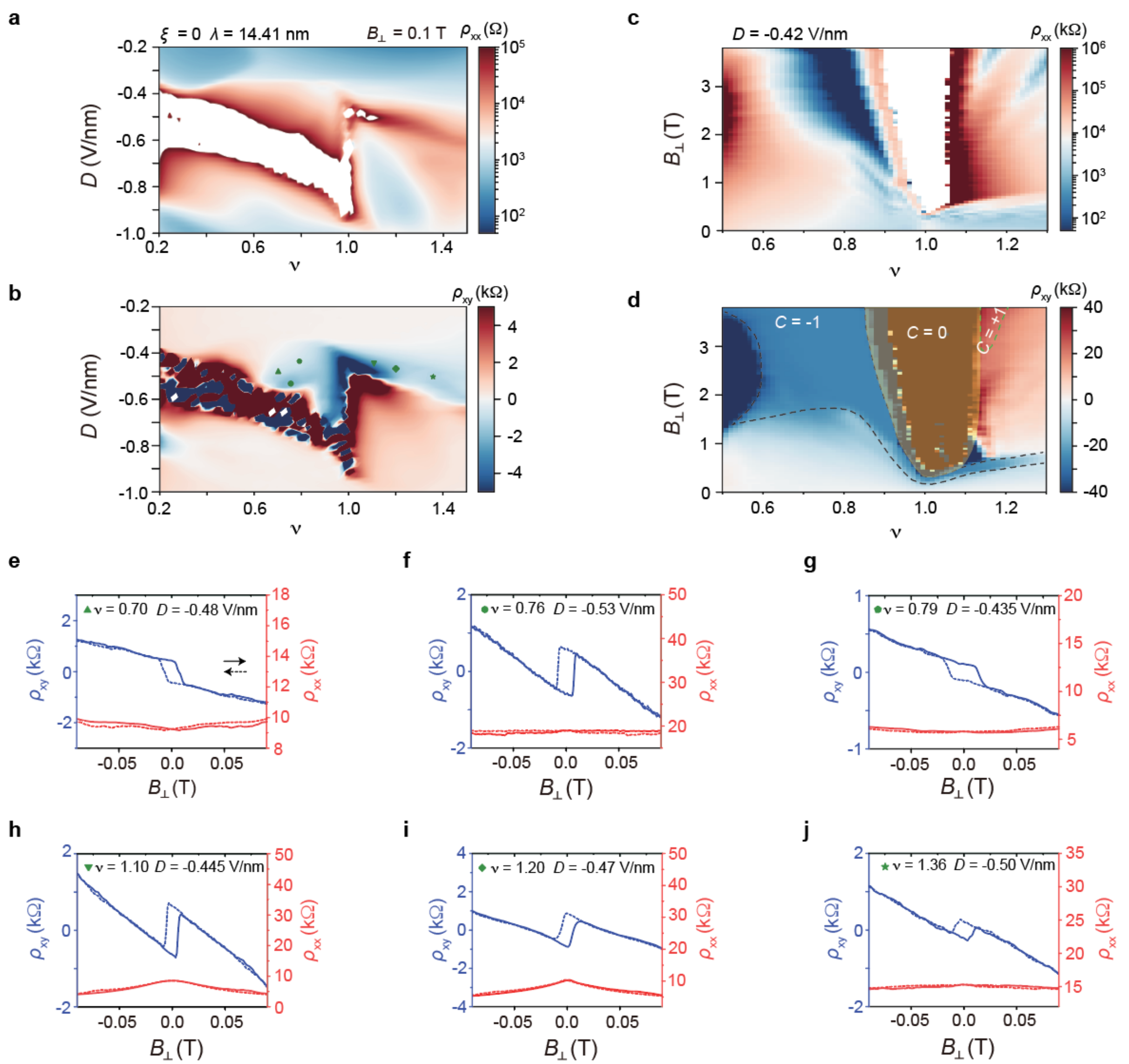

**Extended Data Fig. 8 The AH effect in the moiré-proximal region of device 4 ($\xi$ = 0 alignment). a,b,** Symmetrized $\rho_{xx}$ (**a**) and anti-symmetrized $\rho_{xy}$ (**b**) as a function of $D$ and $\nu$ in device 4, measured at $B_\perp$ = ±0.1 T and $T$ = 10 mK. **c**,**d,** $\rho_{xx}$ (**c**) and $\rho_{xy}$ (**d**) as a function of $B_\perp$ and $\nu$ at $T$ = 10 mK, with fixing $D$ = -0.42 V/nm. Chern insulators with $C$ = -1 emerge under a small $B_\perp$, exhibiting a similar curved trajectory in $\nu$- $B_\perp$ map as observed in device 7 (shown in Figs. 4**d**-4**e**). **e**-**j**, Representative magnetic hysteresis loops of $\rho_{xx}$ and $\rho_{xy}$ measured at $T$ = 10 mK, with the corresponding ($\nu$, $D$) points marked in **b**. The AH effect is obviously observed near $\nu$ = 1 in the moiré-proximal region, indicating the spontaneous time-reversal symmetry breaking.

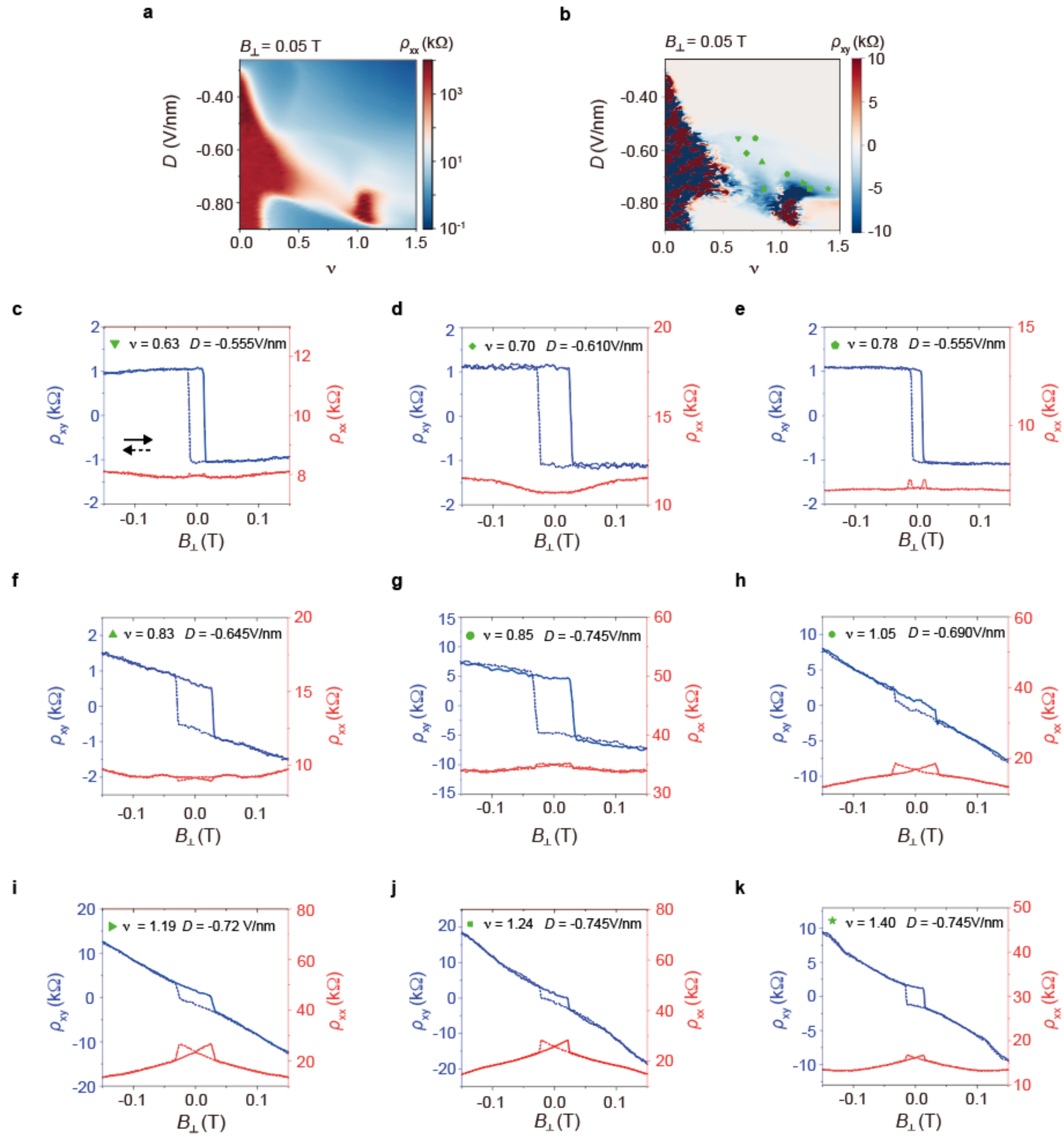

**Extended Data Fig. 9 The AH effect in the moiré-proximal region of device 7 (RPG/hBN, $\xi$ = 0 alignment)**. **a,b,** Symmetrized $\rho_{xx}$ (**a**) and anti-symmetrized $\rho_{xy}$ (**b**) as a function of $D$ and $\nu$ in device 7, measured at $B_\perp = \pm 0.05$ T and $T = 20$ mK. **c**-**k**, Representative magnetic hysteresis loops of $\rho_{xx}$ and $\rho_{xy}$ measured at $T = 20$ mK, with the corresponding ($\nu$, $D$) points marked in **b**. The AH effect emerges within a certain range of $D$ and $\nu$ near $\nu = 1$ in the moiré-proximal regime.

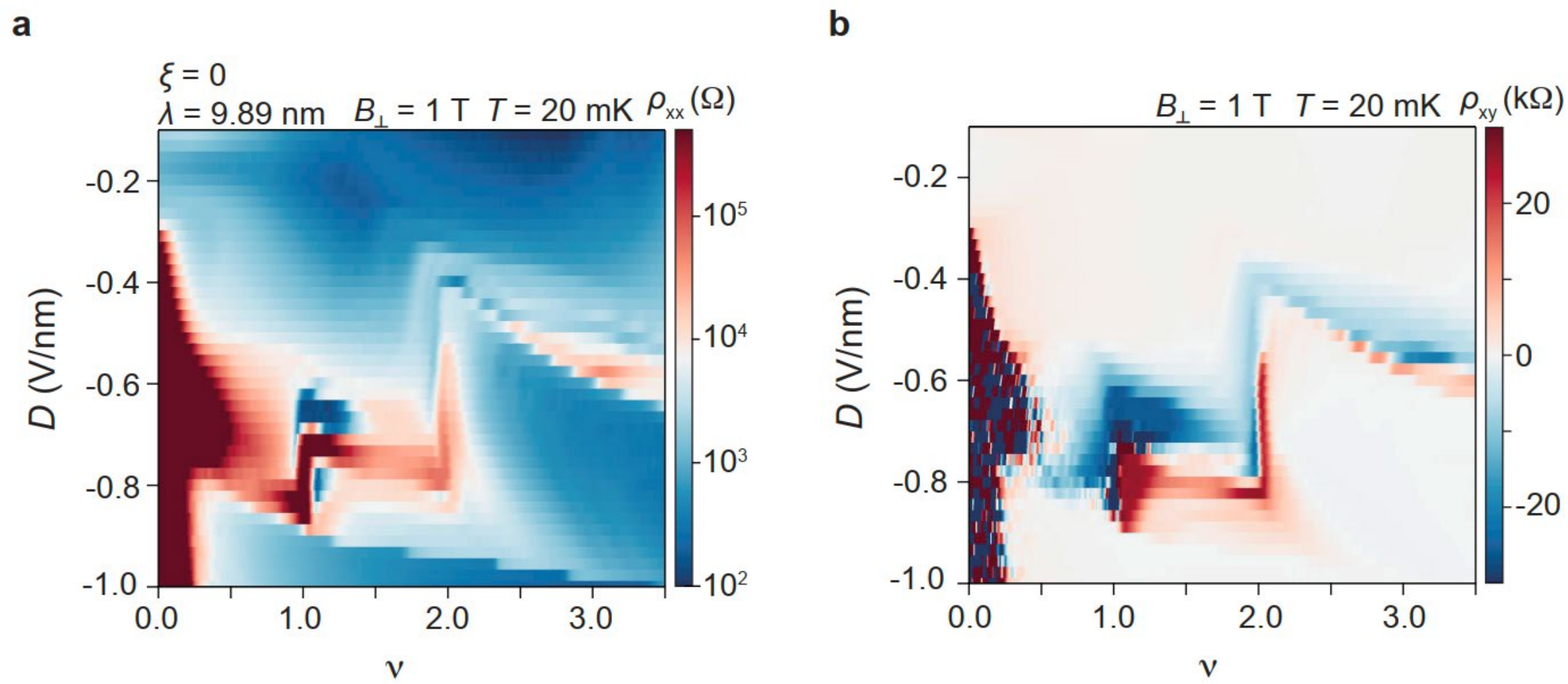

**Extended Data Fig. 10 The *ν*-*D* phase diagram at $B_\perp$ = 1 T in the moiré-proximal region of device 7 (*ξ* = 0 alignment)**. **a**, **b**, Symmetrized $\rho_{xx}$ (**a**) and anti-symmetrized $\rho_{xy}$ (**b**) maps as functions of *D* and *ν* in device 7, measured at $B_\perp = \pm 1$ T and $T$ = 20 mK. Notably, at $\nu = 1$, a trivial insulator emerges within the *D*-field ranges about from -0.9 V/nm to -0.68V/nm. The insulating state is flanked by Chern insulators with $C = +1$ and $C = -1$. The ICI appears only within a narrow range of *D* about -0.9 V/nm to -0.55V/nm, underscoring the sensitivity of the ICI states near $\nu = 1$ to the applied *D* in the moiré-proximal region.